\documentclass[12pt,epsf,amssymb]{article}

\usepackage{graphicx}
\usepackage{axodraw}
\usepackage{amsfonts}      
\usepackage[usenames]{color}
\usepackage{amsmath}

\begin{document}
\baselineskip 0.6cm
\newcommand{\gsim}{ \mathop{}_{\textstyle \sim}^{\textstyle >} }
\newcommand{\lsim}{ \mathop{}_{\textstyle \sim}^{\textstyle <} }
\newcommand{\vev}[1]{ \left\langle {#1} \right\rangle }
\newcommand{\bra}[1]{ \langle {#1} | }
\newcommand{\ket}[1]{ | {#1} \rangle }
\newcommand{\Dsl}{\mbox{\ooalign{\hfil/\hfil\crcr$D$}}}
\newcommand{\nequiv}{\mbox{\ooalign{\hfil/\hfil\crcr$\equiv$}}}
\newcommand{\nsupset}{\mbox{\ooalign{\hfil/\hfil\crcr$\supset$}}}
\newcommand{\nni}{\mbox{\ooalign{\hfil/\hfil\crcr$\ni$}}}
\newcommand{\EV}{ {\rm eV} }
\newcommand{\KEV}{ {\rm keV} }
\newcommand{\MEV}{ {\rm MeV} }
\newcommand{\GEV}{ {\rm GeV} }
\newcommand{\TEV}{ {\rm TeV} }

\def\diag{\mathop{\rm diag}\nolimits}
\def\tr{\mathop{\rm tr}}

\def\Spin{\mathop{\rm Spin}}
\def\SO{\mathop{\rm SO}}
\def\O{\mathop{\rm O}}
\def\SU{\mathop{\rm SU}}
\def\U{\mathop{\rm U}}
\def\Sp{\mathop{\rm Sp}}
\def\SL{\mathop{\rm SL}}
\def\simgt{\mathrel{\lower2.5pt\vbox{\lineskip=0pt\baselineskip=0pt
           \hbox{$>$}\hbox{$\sim$}}}}
\def\simlt{\mathrel{\lower2.5pt\vbox{\lineskip=0pt\baselineskip=0pt
           \hbox{$<$}\hbox{$\sim$}}}}

\def\change#1#2{{\color{blue} #1}{\color{red} [#2]}\color{black}\hbox{}}


\begin{titlepage}

\begin{flushright}
UCB-PTH-06/15; LBNL-61344 \\
\end{flushright}

\vskip 0.75cm
\begin{center}
{\large \bf Landscape Predictions for the Higgs Boson and Top Quark Masses} 

\vskip 1.2cm
Brian Feldstein, Lawrence J. Hall and Taizan Watari\footnote{address
 after August `06: Caltech, Pasadena, CA 91125, USA}

\vskip 0.4cm
{\it Department of Physics and Lawrence Berkeley National 
Laboratory,

University of California, Berkeley, CA 94720, USA} \\

\vskip 1.25cm

\abstract{If the Standard Model is valid up to scales near the Planck 
mass, and if the cosmological constant and Higgs mass parameters scan 
on a landscape of vacua, it is well-known that the observed orders 
of magnitude of these quantities can be understood from environmental 
selection for large scale structure and atoms. 
If in addition the Higgs quartic coupling scans, with a probability 
distribution peaked at low values, environmental selection 
for a phase having a scale of electroweak symmetry breaking much less 
than the Planck scale leads to a most probable Higgs mass of $106 \pm 6$
 GeV for $m_{t} = 171 \pm 2$ GeV. 
While fluctuations below this are negligible, the upward fluctuation 
is $25/p$ GeV, where $p$ measures the strength of the peaking of the a
priori distribution of the quartic coupling. 
There is an additional $\pm 6$ GeV uncertainty from calculable higher 
loop effects, and also sensitivity to the experimental value of $\alpha_s$.  If the top Yukawa coupling also scans, 
the most probable top quark mass is predicted to lie in the range 
(174---178) GeV, providing the standard model is valid to at least
$10^{17}$ GeV, with an additional uncertainty of $\pm 3$ GeV from higher loops.
The downward fluctuation is 35 GeV/$\sqrt{p}$, suggesting that $p$ is 
sufficiently large to give a very precise Higgs mass prediction.
While a high reheat temperature after inflation could raise 
the most probable value of the Higgs mass to 118 GeV, 
maintaining the successful top prediction suggests that reheating 
is limited to about $10^8$ GeV, and that the most probable value 
of the Higgs mass remains at 106 GeV. If all Yukawa couplings scan, 
then the $e,u,d$ and $t$ masses are understood to be outliers 
having extreme values induced by the pressures of strong environmental
selection, while the  $s, \mu, c, b, \tau$ Yukawa couplings span
only two orders of magnitude, reflecting an a priori distribution
peaked around $10^{-3}$.
An interesting extension to neutrino masses and leptogenesis follows 
if right-handed neutrino masses scan, with a preference for larger values, 
and if $T_R$ and $T_{\rm max}$ scan with mild distributions.  
The broad order of magnitude of the light neutrino masses and 
the baryon asymmetry are correctly predicted, 
while the right-handed neutrino masses, the reheat temperature 
and the maximum temperature are all predicted to be of order $10^8\mbox{--}10^9$ GeV.}

\end{center}
\end{titlepage}


\section{Introduction}
The Standard Model (SM) of particle physics
is both extremely successful 
and highly predictive.  It has passed successive hurdles,
from the discovery of weak neutral currents to the precision
electroweak data from $10^7$ Z decays. It even explains why protons
are so stable and why neutrinos are so light.
Despite these successes, it is generally not
viewed as a fundamental theory, but as an effective field theory valid
on scales less than about a TeV.  This is because the
Higgs boson mass parameter receives radiative corrections that are
quadratically divergent, and therefore proportional to
$\Lambda_{SM}^2$, where $\Lambda_{SM}$ is the maximum mass scale that the
theory describes.  For large values of $\Lambda_{SM}$, tree-level and
radiative contributions to the Higgs mass parameter must cancel
to a fractional precision of
\begin{equation}
\Delta \approx \left( \frac{0.5\, \mbox{TeV}}{\Lambda_{SM}}\right)^2
 \left( \frac{m_H}{130 \, \mbox{GeV}}\right)^2,
\label{eq:finetune}
\end{equation}
where $m_H$ is the physical Higgs boson mass.
For the SM to be valid up to 5 TeV, a 
cancellation by two orders of magnitude is already required, and to reach the Planck scale
requires an adjustment finely tuned to 32 orders of magnitude.  Theories that solve this
naturalness problem, including technicolor, supersymmetry, composite Higgs
bosons and extra spatial dimensions, have almost defined physics
beyond the SM,  and are the main focus of potential discoveries at the
Large Hadron Collider.

Rather than just being a description of
interactions beneath the TeV scale,
if the SM is valid up to very large energies, for example to the
Planck scale $M_{Pl} \simeq 2.4 \times 10^{18} \; \GEV$,  then it
predicts the mass of the Higgs boson to be in the range
\begin{equation}
106 \, \mbox{GeV} < m_H < 180 \, \mbox{GeV}.
\label{eq:mHpred}
\end{equation}
The lower limit arises from stability of the SM vacuum
\cite{vacstab,vacstab2,vacstab3,Strumia}. 
A light Higgs boson results in the quartic Higgs self interaction becoming
negative at large energies; if it is too negative, then the universe
undergoes quantum tunneling to a phase quite unlike the observed
phase.  On the other hand the upper bound results from
perturbativity: if the Higgs mass is too heavy, the quartic self
coupling becomes non-perturbative at scales well below the Planck
scale. It is remarkable that the prediction (\ref{eq:mHpred}) receives so
little attention, since it closely coincides with the current
experimental 95\% C.L. range 
\begin{equation}
114 \, \mbox{GeV} < m_H < 175 \, \mbox{GeV}
\label{eq:mHdata}
\end{equation}
from direct searches and precision electroweak data,
respectively \cite{mHdata}.
For a natural SM with $\Lambda_{SM} \approx$ TeV, the Higgs quartic
coupling could lie in a range spanning two orders of magnitude,
0.03---3,  whereas for a large value of $\Lambda_{SM}$ the range is
narrowed to a factor of 3, and consequently $m_H$ is constrained 
to a factor of about $\sqrt{3}$.

If the LHC discovers physics beyond the SM, or if it discovers a heavy
Higgs boson, then 
the apparent agreement between (\ref{eq:mHpred}) and (\ref{eq:mHdata}) 
will be seen to be a coincidence.
However, if the LHC discovers a SM Higgs boson in the range of 
(\ref{eq:mHpred}), and no physics beyond the SM, then the case 
for a very large $\Lambda_{SM}$ will be considerably strengthened.  
In this case the LHC would have verified and extended 
the Little Hierarchy Problem already visible at LEP \cite{LHP}, and 
there would be two obvious interpretations:  new physics could appear
at several TeV and the SM could be accidentally unnatural with
apparent fine-tunings at the percent level; alternatively the
conventional understanding of naturalness could be completely wrong and
$\Lambda_{SM}$ could be extraordinarily large. 
The LHC would eventually yield a very precise measurement of the Higgs
boson mass; could this discriminate between these two interpretations?

An alternative to naturalness is the idea that the universe contains many 
patches with differing underlying physics, and only those patches with a 
certain complexity will be the subject of observation. Thus certain 
observations can be explained from the selection of complexity rather than 
from symmetry principles \cite{Hoyle}.
If the weak scale is selected anthropically, i.e. by environmental 
requirements for complexity, the concept of naturalness is not needed to understand the 
hierarchy between weak and Planck scales \cite{Agrawal:1997gf}.  In this picture, the
fundamental theory of nature may contain a huge number of vacua, and the
Higgs mass parameter may depend on the vacuum. If the universe contains many
patches, each with its own vacuum, then essentially all possible
values of the weak scale are realized somewhere in the universe. It is
only in those patches where the weak scale is in the range of 100s of
GeV that atomic physics provides the building blocks for carbon based
life. The cut-off scale of the SM, $\Lambda_{SM}$, may be extraordinarily 
large, and the Higgs expectation value of most vacua may be of order of 
$\Lambda_{SM}$, yet such vacua are irrelevant to us, since we necessarily 
find ourselves in one of the very rare patches having hospitable chemistry.  

Despite resistance from some physicists, these ideas---a landscape of vacua 
and environmental selection---have gained attention over the last decade:
selection of patches
of the universe containing large scale structures, such as galaxies, can
essentially solve the cosmological constant problem and
predicts that observers are likely to inhabit patches containing dark energy  
\cite{weinberg}.
Furthermore, the number of known vacua of string theory continues to increase, 
and the string landscape appears to be able to scan parameters 
sufficiently densely to allow environmental selection \cite{stringlandscape}.
Such ideas should perhaps be resisted, since they are so difficult to test 
experimentally. How are we to test such theories when the other patches 
of the universe lie outside our horizon?  
Natural theories are tested by assuming a theory with a parameter space
that is sufficiently restricted that predictions for observables can
be made.  In the landscape, predictions may be possible by combining environmental 
selection with assumptions about the underlying vacuum probability
distributions.
In this paper we argue that the Higgs boson mass is well suited to this sort of
prediction, since it depends on the quartic scalar coupling, and if
this coupling is too low the SM vacuum decays.  If the landscape
favors low values of this coupling, environmental selection will
favor Higgs masses close to the stability bound. 

\begin{figure}[t]
\begin{center}
\includegraphics[width=0.5\linewidth]{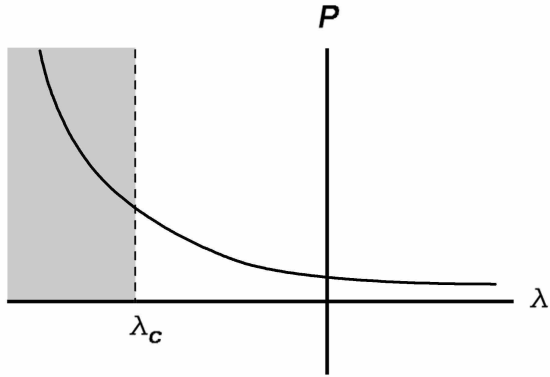}
\caption{\label{fig:P} The a priori probability distribution is peaked at low 
values of $\lambda$. }
\end{center}
\end{figure}

Expanding the Higgs potential $V(\phi)$ in powers of $\phi/\Lambda_{SM}$, 
the relevant part of the Higgs potential is 
\begin{equation}
V(\phi) = \Lambda^4 + m^2 \phi^\dagger \phi + \lambda (\phi^\dagger \phi)^2.
\label{eq:V}
\end{equation}
In this paper we assume that the entire SM Higgs potential varies, or scans, 
from one patch of the universe to another.
The scanning of $\Lambda^4$ leads to the scanning of the
cosmological constant and allows an environmental selection for galaxies, largely explaining 
the observed value of the dark energy \cite{weinberg}. 
Similarly, the scanning of the Higgs mass parameter $m^2$ and
quartic coupling $\lambda$ allows 
for the environmental selection of required atomic properties, determining 
the weak scale $\sqrt{- m^2/\lambda}$ \cite{Agrawal:1997gf}.
In those regions of the universe selected environmentally by  both the cosmological constant
and the weak scale, there will be a variation in the coupling
$\lambda$ and therefore in the Higgs boson mass from one patch to another. 
What are the physical consequences of this scanning of the Higgs quartic coupling
$\lambda$?  Environmental selection will choose the required
electroweak phase: we must live in a region with $\lambda$ above
a minimum value, $\lambda_c$.  This is how the usual stability limit on the SM Higgs
boson mass arises in the landscape, leading to the lower bound
in (\ref{eq:mHpred}).  For $\lambda > \lambda_c$ there is no obvious
environmental selection of one value of $m_H$ over another, hence the
prediction for the Higgs boson mass is governed by the
probability distribution\footnote{To be more precise, by the probability
distribution $P(m^2,\lambda)|_{\sqrt{-m^2/\lambda}=174\,
\GEV}$. Hereafter, we mean this by $P(\lambda)$.} 
$P(\lambda)$ of the coupling $\lambda$ in the vacua with acceptable
large scale structure and atomic physics.  To proceed, apparently one
needs a calculation of $P(\lambda)$ from the landscape. In fact, only a single
assumption on its form is required: we assume that $P(\lambda)$ is sufficiently 
peaked at low values that $\lambda$ is expected to be
near $\lambda_c$, as illustrated in Figure \ref{fig:P}, corresponding to a Higgs
mass near the stability bound. 
The precision of this prediction
clearly depends on how steep $P(\lambda)$ is near $\lambda_c$, as we
investigate in some detail.


It is important to stress that our prediction does depend on an
assumption for $P(\lambda)$.  For example, an alternative
assumption that $P(\lambda)$ is peaked at large positive values of
$\lambda$ predicts $m_H$ near the perturbativity limit of 180 GeV.  
However, this prediction could depend on whether the theory 
has a large energy interval in which it is
strongly coupled before a perturbative quartic emerges at low energy.  
Furthermore, an observation of a Higgs boson mass in this region could be
interpreted in terms of a  strongly coupled fundamental theory without
any recourse to the landscape.  
In the rest of this paper, we concentrate on the
possibility that the probability distribution $P(\lambda)$ is peaked
towards low values, corresponding to a Higgs boson mass near the
stability limit of $m_H \approx 106$ GeV. A measurement of such a Higgs
mass, together with the absence of any new physics, would provide
evidence for environmental selection.

The top quark Yukawa coupling, $h$, plays an essential role in electroweak
symmetry breaking in the SM with large $\Lambda_{SM}$, through its
effect on the renormalization group (RG) scaling of the Higgs quartic
coupling, $\lambda$.  Suppose that $h$ also scans in the landscape,
so that there is a combined probability distribution $P(\lambda, h)$.
How is the discussion of the Higgs mass prediction changed?  
As we will show, the metastability boundary in the $(\lambda, h)$
plane has a special point, where $\lambda$ is at a minimum, 
and a simple assumption for the a priori probability distribution, 
$P(\lambda, h)$, implies that patches of the universe 
in the desired electroweak phase are most likely to be 
in the neighborhood of this point.
Remarkably, at this special point 
\begin{equation}
m_{H_c} = (121 \pm 6) \, \GEV \hspace{1in}
m_{t_c} = (176 \pm 2) \,  \GEV,
\label{eq:crit}
\end{equation}
where the uncertainty corresponds to $\Lambda_{SM} = 10^{18 \pm
1}$ GeV,
leading simultaneously to a prediction for the Higgs mass and to a broadly
successful post-diction for the top mass. 
Although there are further uncertainties of about $\pm 6$ GeV and $\pm
3$ GeV on $m_{H_c}$ and $m_{t_c}$ respectively, from higher loop
effects, this result is nevertheless very striking.\footnote{We should
mention here that   
predictions for the Higgs and top masses in the SM, assumed valid to
very high scales, have been made based on very different principles.  
One prediction resulted from assuming that there are two degenerate 
electroweak vacua, with one occuring near the Planck scale \cite{FN},  
while another resulted from the assumption that two phases should
coexist, giving borderline vacuum metastability \cite{FNT}.}

It is well-known that the stability limit on $m_H$ in the SM
depends on $T_R$, the reheat temperature of the
universe after inflation.  
In the landscape, $T_R$ may scan from one patch of the universe 
to another, or it might be fixed; either way, we study the landscape 
Higgs mass prediction in two cases.  In the first case, 
in sections 2 and 3, we assume that the probability distribution 
is dominated by patches having $T_R \simlt 10^8$ GeV, 
so that thermal fluctuations can be ignored, and the
stability limit arises from quantum fluctuations from the false
vacuum.  In section 2 we keep the top Yukawa coupling fixed, while in
section 3 we allow it to scan. In the second case, in section 4, 
we assume that the landscape probability distribution is dominated
by patches having $T_R \simgt 10^8$ GeV, and 
study the Higgs mass prediction arising from thermal nucleation 
of bubbles of the true vacuum. 
In section 5 we embed these ideas in the ``scanning SM'' where all SM
parameters scan, discussing charged fermion mass hierarchies, the scale
of neutrino masses, leptogenesis and consequences for inflation.  
Conclusions are presented in section 6.

\section{Higgs Mass Prediction from Quantum Tunneling}

\subsection{Quantum Tunneling}

In this section we predict the Higgs boson mass 
using the environmental constraint of sufficient stability of the
electroweak vacuum in a landscape scenario. 
The SM is assumed to be valid up to a high energy scale, $\Lambda_{SM}$, such as the Planck scale. 
To be precise, we define $\Lambda_{SM}$ to be the scale at which the standard 
model RG equations and bounce action calculations  are
still valid to within $1\%$.  Thus $\Lambda_{SM}$ is slightly 
lower than the scale at which new physics actually arises.
We concentrate on the consequences of scanning $(\Lambda^4, m^2, \lambda)$, 
assuming that all other parameters of the SM  do not effectively scan
or are fixed tightly to observed values by other environmental constraints.
In the patches of the universe dominating the probability
distribution,
the maximum temperature after inflation is assumed to be sufficiently low 
($T_R \simlt 10^8\, \GEV$) that thermal nucleation of the phase
transition is sub-dominant.

\begin{figure}[t]
\begin{center}
\begin{tabular}{cc}
\includegraphics[width=.5\linewidth]{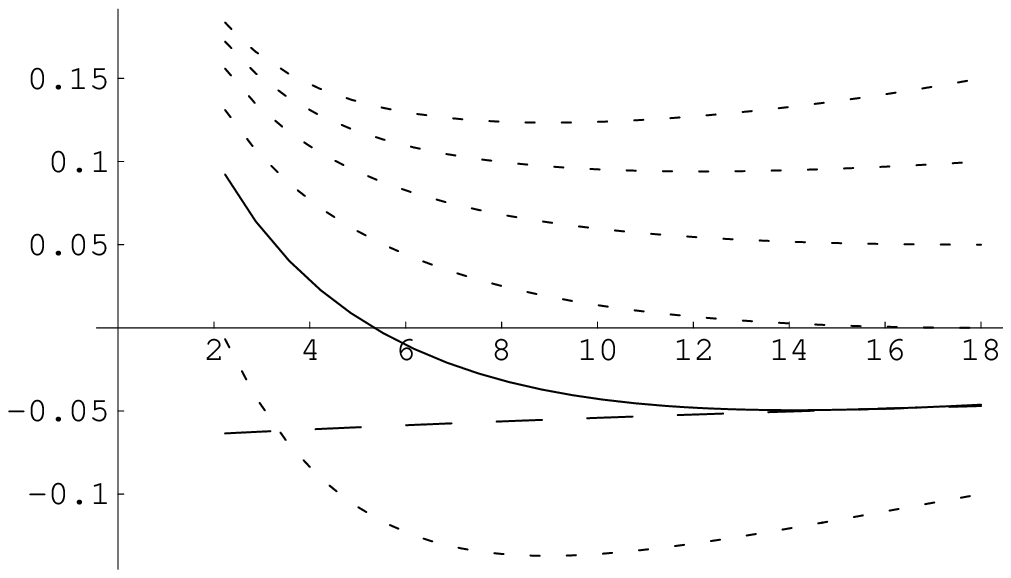} &
\includegraphics[width=.4\linewidth]{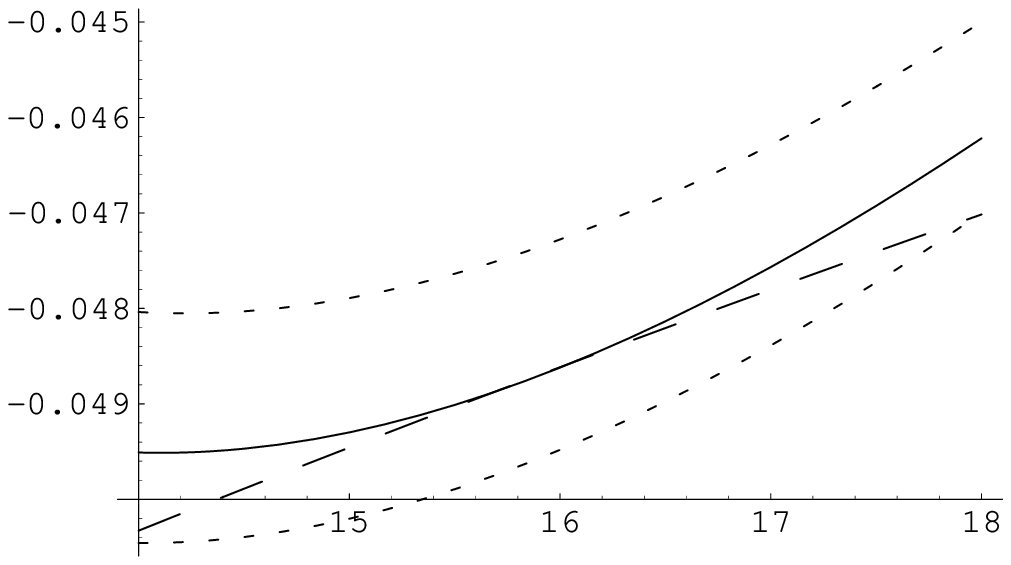} 
\end{tabular}
\begin{picture}(0,0)(450,75)
\Text(15,150)[b]{$\lambda$}
\Text(230,70)[b]{$\log_{10} \mu$}
\Text(430,20)[bl]{$\log_{10} \mu$}
\Text(275,130)[b]{$\lambda$}
\Text(365,145)[b]{$\vev{\rm Blow~up~of~the~figure~on~the~left}$}
\Text(130,40)[t]{$\lambda_c(M)$}
\Text(310,45)[tl]{$\lambda_c(M)$}
\end{picture}
\caption{\label{fig:running4} 
Several RG trajectories for the Higgs quartic coupling, 
$\lambda$, are shown by dotted curves.  
The dashed line gives the critical coupling $\lambda_c$ at each energy, 
defined by (\ref{eq:lambdac}). 
The special trajectory that corresponds to the usual SM metastability
bound is shown by a solid curve.
Any patch of the universe with a trajectory that passes significantly 
into the region below the dashed line is likely to decay out 
of the false vacuum with $\vev{H}=v$ before a cosmological time scale 
of order $10^{10}$ years.   
In this figure, $\Lambda_{SM}$ is chosen to be $10^{18}$ GeV, 
and $(g_Y,g_L,g_s,h)|_{\mu = \Lambda_{SM}}  = 
(0.466,0.512,0.500,0.391)$ yielding, for the special trajectory with 
$\lambda_0(\Lambda_{SM})= - 0.0462$, weak-scale SM parameters 
$(\alpha(m_Z)^{-1},\sin^2 \hat{\theta}_W (m_Z),\alpha_s(m_Z),m_t) = 
 (127.9,0.2313,0.1176,171 \, \GEV)$ \cite{PDG06} 
as well as $m_H = 106$ GeV. 
The analysis is performed with 2-loop RG equations \cite{RGE} and 1-loop 
(2-loop in $g_s$) threshold corrections \cite{threshold1,Sher,threshold3} at the matching
 scale $\mu = m_t$ \cite{Hambaye}.}
\end{center}
\end{figure}

At large field values, the effective potential of the Higgs field 
$H = \sqrt{2} \, {\rm Re} (\phi^0)$ is well approximated by
%
\begin{equation}
V(H) =  \frac{\lambda(H)}{4} H^4 \qquad \qquad {\rm for~}H \gg v \equiv 246 \; \GEV,  
\label{eq:quartic-pot}
\end{equation}
where $\lambda(H)$ is the value of the running Higgs quartic coupling evaluated at the 
scale $\mu = H$.  The RG equation for $\lambda$ has the 1-loop form
\begin{equation}
16\pi^2 \frac{d\lambda}{d \ln \mu} = 24\lambda^2 + 12\lambda h^2 -6 h^4 -9 \lambda g_L^2 
- 3\lambda g_Y^2
+\frac{3}{8}g_Y^4 +\frac{3}{4}g_L^2 g_Y^2 +\frac{9}{8}g_L^4,
\label{eq:lambdarge}
\end{equation}
where $h$ is the top Yukawa coupling, and $g_L$ and $g_Y$ are 
the $SU(2)_L$ and $U(1)_Y$ gauge couplings, respectively.  
Several trajectories for $\lambda(\mu)$ are illustrated in Figure \ref{fig:running4}.
The effect of the large top Yukawa coupling is to reduce the value of $\lambda$ 
at high energies. In fact, $\lambda$ will become negative below $\Lambda_{SM}$ 
if its value at $\mu \approx v$ is too small.  
In such cases, the minimum with $\vev{H} =v$ is not the true minimum 
of the potential, and  we must consider the possibility of the universe 
tunneling quantum mechanically from our hospitable $\vev{H} = v$ vacuum 
to the inhospitable one with very large $\vev{H}$.\footnote{In this phase the ratio
of quark masses to the QCD scale and to the Planck mass are very far 
from allowing conventional nuclei and cosmology.} 
Throughout we assume that conditions of the early universe lead to 
a sufficient number of patches with the desired metastable phase.

The quantum tunneling rate per unit volume to the true vacuum is dominated 
by a bounce solution $H(r)$. A pure up-side-down quartic potential, 
(\ref{eq:quartic-pot}) with negative $\lambda$, has an SO(4) symmetric 
bounce solution \cite{LW}. Due to the conformal nature of the potential 
(ignoring the running of $\lambda$ and the quadratic term of order the
weak scale), there is a family of bounce solutions, $H(r)=cH_0(cr)$ for
$c>0$, with different size $\propto 1/c$ and field value at the center 
$H(r=0) \propto c$. A bounce solution with $H(r=0) = M$ contributes to
the decay rate an amount 
\begin{equation}
\Gamma(M) \approx  M^4 e^{-\frac{8\pi^2}{3|\lambda(M)|}},
\label{eq:rateM}
\end{equation}
so that the total decay rate is approximated by \cite{vacstab,vacstab2,vacstab3,Strumia}
\begin{equation}
\Gamma[\lambda_0] \approx \displaystyle\max_{M < \Lambda_{SM}} \Gamma(M)
= \displaystyle\max_{M < \Lambda_{SM}}
\left[ M^4 e^{-\frac{8\pi^2}{3|\lambda(M)|}} \right].
\label{eq:rate-vac}
\end{equation}
Each patch of the universe has its own $\lambda_0 \equiv \lambda(\Lambda_{SM})$, 
RG trajectory $\lambda(\mu)$, 
and corresponding decay rate $\Gamma[\lambda_0]$.
By time t, an arbitrary point remains in the desired false vacuum 
with $\vev{H}=v$ provided no bubble nucleated in its past light cone. 
Since the vacuum tunneling rate $\Gamma[\lambda_0]$ does not depend on
time, bubble nucleation is most likely to occur at the epoch with 
time of order $t$, rather than at a much earlier epoch. 
Assuming that the universe has been matter dominated 
most of the time until $t$, the fraction of the volume of a patch 
in the false vacuum at time $t$ is given by \cite{GW}
\begin{equation}
f(\lambda_0; t) =  e^{-\Gamma[\lambda_0]t^4},
\label{eq:GW}
\end{equation}
where a coefficient of order one in the exponent is ignored.\footnote{
The exponent depends non-perturbatively on the Higgs quartic coupling 
$\lambda$ as in (\ref{eq:rateM}), and, if $|\lambda|$ is small, 
a small variation in the value of 
$\lambda$ can change $\Gamma$ by orders of magnitude. 
Thus, any prefactor of order unity to $\Gamma t^4$ is unimportant compared 
with the sensitivity on $\lambda$. 
}$^,$\footnote{As the universe evolves into an era 
dominated by the cosmological constant the fraction of the volume in the 
false vacuum is given asymptotically by 
$f(\lambda_0; t)= e^{-\Gamma[\lambda_0] t/H_0^3}$, where 
$H_0$ is the Hubble parameter and
a coefficient of order one in the exponent is again ignored.}

For practical applications, we are interested in $t$ of order $10^{10}$ years; 
$t$ is either still in the matter-dominated era or in the dark-energy 
dominated era so that the exponent is roughly of order 
$\Gamma[\lambda_0] \tau^4$, where $\tau = 10^{10}$ years. 
If $\Gamma(M) \tau^4 \simgt 1$ for any scale 
in the range $v \ll M \simlt \Lambda_{SM}$, the survival fraction  
$f(\lambda_0; \tau)$ is significantly less than 1. 
Since $\Gamma(M)$ primarily depends on $\lambda(\mu)$ renormalized at 
$\mu = M$, one can introduce a critical value of the quartic coupling 
$\lambda_c(M)$ by $\Gamma(M) \tau^4 \simeq 1$
\begin{equation}
\lambda_c(M)  \simeq    - \frac{2 \pi^2}{3}  
 \left( \frac{1}{\ln (M_{Pl} \tau) + \ln
   \frac{M}{M_{Pl}}} \right) \approx - 0.047 \left(1 + \frac{1}{138}
  \ln \frac{M_{Pl}}{M} \right),
\label{eq:lambdac}
\end{equation}
where $\ln (M_{Pl} \tau) \approx 138$ is used to obtain 
the last expression.
This critical line  of stability, $\lambda_c(M)$, is shown by 
a dashed line in Figure \ref{fig:running4}.  
Whenever a trajectory $\lambda(\mu)$ dips below 
$\lambda_c(\mu)$ at any scale $v \ll \mu \simlt \Lambda_{SM}$, 
bounces with $H(r=0)$ of order that scale would have destabilized 
the desired false vacuum.
The usual metastability bound on the Higgs boson mass in
\cite{vacstab2,vacstab3,Strumia}
corresponds to the trajectory that touches the critical line 
of stability at one scale $\mu = M_{dom} \sim 10^{16} \, \GEV$, while 
maintaining $\lambda(\mu) \geq \lambda_c(\mu)$ for all $\mu$ up to $\Lambda_{SM}$.  
Thus, the bounce of most danger is the one with $H(r=0) = M_{dom}$, and 
the metastability bound is independent of $\Lambda_{SM}$ provided 
$\Lambda_{SM} > M_{dom}$.
Note that there is an assumption here: for all trajectories of interest, 
the contribution to $\Gamma[\lambda_0]$ from quantum fluctuations to field 
values larger than $\Lambda_{SM}$ are sub-dominant.  We must assume this 
because we cannot calculate these contributions without knowing some of 
the details of the more complete theory above $\Lambda_{SM}$.  
The trajectories of interest include the special one, shown as a solid
curve in Figure \ref{fig:running4}, and all those that lie above it.  
The assumption is reasonable because, as the SM begins to break down 
at energies just above $\Lambda_{SM}$, these trajectories 
all lie above the critical curve $\lambda_c(\mu)$, shown dashed in 
Figure \ref{fig:running4}.

A landscape will provide some a priori probability $P(\lambda_0)$, and 
we follow the principle that the probability distribution of observable 
parameters is further weighted by the fraction of observers who see them.
For patches of the universe having $\lambda(\Lambda_{SM})=\lambda_0$,
only a volume fraction $f(\lambda_0; t)$  remains in our 
desired false vacuum with $\vev{H}=v$, introducing an 
additional $\lambda_0$ dependence in the total probability distribution
\begin{equation}
{\cal P}(\lambda_0;t) d \lambda_0 = 
P(\lambda_0) e^{-\Gamma[\lambda_0] t^4} d \lambda_0.
\label{eq:distr-civil}
\end{equation}
The a priori distribution $P(\lambda_0)$ describes all $\lambda_0$ 
dependence that originates from cosmology up to the end of 
(the last) inflation. Since the Higgs boson mass itself is not environmentally 
important, $f$ will be the only $\lambda_0$ dependence 
(and Higgs boson mass dependence) that originates after the end of inflation. 

One could also study an all-time probability distribution, 
instead of the distribution for contemporary observers.
Such a distribution is obtained by integrating (\ref{eq:distr-civil}) 
over $t$, weighted by the time evolution of the number of observers, $\rho(t)$, 
which we expect to be independent of $\lambda_0$. 
Thus, $f$ still remains the only source of the late-time 
dependence of the distribution on the Higgs boson mass.

\subsection{The Prediction}

Using the probability distribution of (\ref{eq:distr-civil}) we would
like to answer three key questions.
\begin{enumerate}
\item   What is the most probable value of $\lambda_0$, and therefore the Higgs
 mass, as seen by observers, such as us,  living at the present age of
 the universe $t = t_0\sim 1.4 \times 10^{10}$ years?
\item What range of Higgs masses correspond to a probability within a factor of, 
say, $1/e$ of the peak probability?  
\item What fraction of observers in the ``multiverse'' live at the time $t_0$ 
or later?
\end{enumerate}
The answer to the first question gives the central value of the Higgs
mass prediction, while the answer to the second gives a measure of the
uncertainty of the prediction.  We will be particularly interested in
how these two answers depend on the form of the a priori probability
distribution $P(\lambda_0)$.  We suspect that a precise prediction
will follow only if there is some peaking in this distribution. 
There may be a trade-off: if the distribution is too weak, 
then we cannot predict the Higgs mass with significant accuracy;  
if it is too strong, then we may find ourselves living in an unstable 
patch of the universe that was extremely lucky to survive until now.  
With the third question, 
we thus investigate how much peaking we can tolerate before observers
living as late as $t_0$ become a rarity.

We assume that $\Lambda_{SM} \simgt M_{dom}$. Since tunneling is 
most likely to occur at the scale $M_{dom}$,  
as can be seen from figure  \ref{fig:running4}, with this assumption the analysis 
becomes independent of $\Lambda_{SM}$.   Consider patches of
the universe with various  values of $\lambda(M_{dom})$ as illustrated 
by the trajectories in Figure  \ref{fig:running4}. An a priori probability distribution
peaked at low values of $\lambda_0$, as illustrated in  Figure
\ref{fig:P}, means that there are more patches with $\lambda(M_{dom})$ below
the critical value $\lambda_c(M_{dom})$ than above.  However, these patches are
more likely to decay. The most probable observed value of $\lambda$ today is
therefore determined by a competition between $P(\lambda_0)$ and
$e^{-\Gamma t_0^4}$ in (\ref{eq:distr-civil}), and is given
by\footnote{We ignore the difference between $t_0$ and $\tau=10^{10}$
years, and use the latter as a reference. $\overline{m}_H$ in
(\ref{eq:pred-val}) increases no more than 0.1 GeV by using $t_0^4$ 
instead (e.g., \cite{Strumia}).}
\begin{equation}
 \overline{\lambda}(M_{dom}) = \lambda_c(M_{dom}) \left( 1 - \frac{1}{540} \ln \frac{540}{p}
\right)
\label{eq:peaklambda}
\end{equation}
where the parameter $p$ describes the strength of the peaking of the a priori probability
distribution at $\lambda_0 = \lambda_c(\Lambda_{SM}) \equiv \lambda_{c,0}$ and is defined by
\begin{equation}
 p = \left. \frac{\partial \ln P(\lambda_0)}{ \partial \ln \lambda_0} 
\right|_{\lambda_0 = \lambda_{0,c}}.
\label{eq:p}
\end{equation}
Since $\lambda_c$ is negative, we are interested in positive
values for $p$.  
(Note that we use $\lambda_{0,c}$ and $\lambda_{c,0}$ interchangeably.)
Clearly, except for extremely large
values of $p$, the double exponential form of the decay probability factor
is so powerful that we can just take $\overline{\lambda}(M_{dom}) = 
\lambda_c(M_{dom})$.
Thus the most probable Higgs mass measured by observers at time $t_0$
corresponds to the solid line trajectory in Figure \ref{fig:running4},
which also corresponds to the usual SM metastability
bound.  We find that this trajectory corresponds to a Higgs boson mass\footnote{ 
These results are in agreement with the metastability bound found in 
\cite{Strumia}.  The results in previous versions of our paper gave values of
$\overline{m}_H$ about 5 GeV larger (for a given $m_t$) due to incorrectly neglecting 
part of the top threshold in $h_t$.  We thank G. Giudice for a very useful e-mail}
\begin{equation}
\overline{m}_H =  106 \, \GEV  + 6 \, \GEV  \left( \frac{m_t - 171 \, \GEV}{2 \, \GEV}
\right) - 2.6 \, \GEV \left(\frac{\alpha_s - 0.1176}{0.002}\right)
\pm 6 \, \GEV . 
\label{eq:pred-val}
\end{equation}
There are several theoretical 
uncertainties that contribute to the $\pm 6$ GeV uncertainty quoted in
(\ref{eq:pred-val}), in particular in relating the physical Higgs boson 
mass to the running quartic coupling evaluated at very high energies and in 
relating the physical top quark mass to the running top quark Yukawa coupling.
The present 6 GeV uncertainty in 
$\overline{m}_H$ coming from the $\pm 2$ GeV experimental uncertainty 
in $m_t$ will be reduced by a factor of 2--4 at the LHC, 
depending on how well systematic uncertainties can be understood, 
while a reduction in the uncertainty in $\alpha_s$ may have to wait for the ILC.
Finally, while $\overline{m}_H$ is the 
most probable value for us to observe, a crucial question is the width 
of the probability distribution for the Higgs mass, which will depend 
on the strength $p$ of the peaking of the a priori distribution 
$P(\lambda_0)$.

Reference \cite{ADK} has studied extensions of the SM where
$m^2$ scans but not $\lambda$. It was assumed that $\lambda_0$ is negative
so that the RG trajectory for $\lambda$ passes through zero at some scale
$\Lambda_{cross}$. Only those patches of the universe with $m^2$
negative and $|m^2| < \Lambda_{cross}^2$ will have a metastable
hospitable electroweak phase, and hence environmental selection will
lead to the weak scale being close to $\Lambda_{cross}$, giving an
alternative understanding of a low weak scale.  This is a very 
interesting idea, but we stress that it is very different from the scanning
of $\lambda$ that this paper is based on.  
From Figure \ref{fig:running4} it is
clear that their idea  cannot work in the SM, hence in addition to a light
Higgs boson, they predict other heavy states at the weak scale.

\subsection{The Width of the Prediction}

In the SM the Higgs mass
must lie in the experimentally allowed range of (114--175) GeV.  To see evidence for a landscape, it is
crucial that the probability distribution for the Higgs boson mass is sharply 
peaked compared to this allowed range. How strong an assumption does this 
require for the a priori probability distribution? 

To study this
question we Taylor expand the probability ${\cal P}(\lambda_0) \propto P
e^{-\Gamma \tau^4}$ of
(\ref{eq:distr-civil}) about the most probable value of the quartic
coupling $\overline{\lambda}_0$, and find the couplings $\lambda_\pm$ where  
${\cal P}$ falls to 1/e of its maximum value. We choose $\lambda_- <
\lambda_+$, so that $\lambda_+$ corresponds to the heaviest Higgs
boson that we are likely to observe. The maximum value of the exponent
$\overline{\Gamma} \tau^4$ is $p/540$, which, for reasons discussed in the
next sub-section, we take to be less than unity. It follows that the curve
${\cal P}(\lambda_0)$ is highly asymmetrical about $\overline{\lambda}_0$,
with $\lambda_-$ determined by the very rapid drop off of $e^{-\Gamma \tau^4}$ 
\begin{equation}
\lambda_- = \overline{\lambda}_0 \left( 1 + \frac{1}{540} \ln \frac{540}{p}
\right),
\label{eq:lambda-}
\end{equation}
and $\lambda_+$ determined\footnote{\label{fn:largep}We caution the reader 
that on Taylor expanding $P$ about
$\overline{\lambda}_0$, the expansion is on the verge of breaking down on
reaching $\lambda_+$.   We have checked that (\ref{eq:lambda+}) is
precisely accurate for the cases of power law and exponential
distributions, but in general it should only be taken at the factor of
2 level.}
by the drop off of $P$
\begin{equation}
\lambda_+ = \overline{\lambda}_0 \left( 1 - \frac{1}{p} \right).
\label{eq:lambda+}
\end{equation}

We immediately see that, to obtain a significant Higgs mass
prediction,  there is no need to have a very large $p$; a much
more modestly peaked a priori distribution is sufficient. 
We also see that for modest values of $p$, the
uncertainty in the Higgs mass coming from the landscape, i.e. from the unknown a
priori probability distribution and from the probabilistic nature of
quantum tunneling, is dominated by the $1/p$ contribution to
$\lambda_+$, which {\it increases} the Higgs boson mass above $\overline{m}_H$. 
The difference between $\lambda_-$ and $\overline{\lambda}_0$ corresponds
to a change in the Higgs mass that is negligible compared to the
uncertainties of (\ref{eq:pred-val}).
Hence, while $\overline{m}_H$ of (\ref{eq:pred-val}) is the most probable
value for the Higgs mass, the expected range is from $\overline{m}_H$ up to
$\overline{m}_H + \Delta m_H$ where $\Delta m_H$ is determined from the shift in the 
quartic coupling at $\Lambda_{SM}$:  $\Delta \lambda \approx |\lambda_{0,c}|/p \approx -.05/p$.
RG running will not change this shift by much as it is taken to the weak scale, and since 
$m_H \sim 350 \sqrt{\lambda} \, \GEV$,  we find \footnote{A different expression is required 
for $p \simgt {\cal O}(500)$, but for such large values, the predicted range is so narrow 
that the precise width is unimportant.}
\begin{equation}
\Delta m_H \approx \frac{25 \, \GEV}{p}. 
\label{eq:DeltamH}
\end{equation}
A very precise prediction follows if $p \geq 10$, while a reasonably 
precise prediction requires only $p \geq 3$.
In the more complete theory at the scale of new physics, 
the description of the
relevant landscape of vacua may involve a probability distribution with
some other set of couplings.  The
simple peaking of $P$, corresponding to $p \geq 3$, may have a
simple origin in the distributions of the more complete
theory, because such a peaking is not destroyed by mild RG scaling.

\subsection{The Fraction of Observers Living After $t_0$}

There is a second important consequence of having the a priori
probability distribution strongly peaked at low values of $\lambda_0$:
it becomes more probable for observers in the multiverse to live
soon after non-linearities in the large scale structure appear.
Once again, this can be understood by studying figure
\ref{fig:running4}. As lower values of $\lambda_0$ become more strongly
preferred, so more patches of the multiverse have 
$\lambda(M_{dom}) < \lambda_c(M_{dom})$, 
and these patches typically decay before time $t_0$.  
This is clearly a problem if there are so few patches of the universe
that survive until $t_0$, that we are very rare observers.  How does
this limit the degree of peaking of $P(\lambda_0)$?

To get a feel for this we can study how $P$ varies as $\lambda_0$ is
decreased below $\lambda_c$.  Instead of using $\lambda_0$ as the
variable describing each patch we can use $t$, the time at which the
patch with coupling $\lambda_0$ typically decays. Taylor expanding
about $t_0$ to first order, we find
\begin{equation}
 P(t) = P(t_0) \left( 1+ \frac{p}{140} \ln \frac{t_0}{t} \right).
\label{eq:Pt}
\end{equation}
The earliest relevant time is the time that suitable structures first
went non-linear, $t_{NL}$, and we expect that $\ln (t_0/t_{NL})$ is no
more than a few.  Hence we see that observers at $t_0$ are common in
the multiverse providing the peaking of the probability distribution
satisfies
\begin{equation}
p \leq  \frac{140}{\ln (t_0/t_{NL})},
\label{eq:plimit}
\end{equation}
which certainly allows for a precise Higgs mass prediction, as can be
seen from (\ref{eq:DeltamH}).

Even for larger values of $p$ the probability for observers in the
multiverse at time $t_0$ does not drop extremely quickly.
Approximating the time evolution of the number of observers in an
undecayed patch, $\rho(t)$, as a step function at $t_{NL}$, the
fraction of observers in the multiverse living at time $t_0$ or later is 
\begin{equation}
f = \frac{I(t_0)}{I(t_{NL})}
\label{eq:f}
\end{equation}
where
\begin{equation}
I(t) = \int_t^\infty P(\lambda_0) e^{- \Gamma[\lambda_0] t'^4} \, dt' d\lambda_0. 
\label{eq:I}
\end{equation}
To evaluate this expression an explicit form for $P$ is required.  For
example, for an exponential distribution
\begin{equation}
P = C e^{p \frac{\lambda_0}{\lambda_c}}
\label{eq:Pexp}
\end{equation}
we find
\begin{equation}
f \approx \left( \frac{t_{NL}}{t_0} \right)^{\frac{p}{140}},
\label{eq:fexp}
\end{equation}
so that the fraction of observers at $t_0$ is not highly suppressed 
even for $p$ as large as several hundred.
For a power law distribution, the times $t$ in (\ref{eq:fexp}) are replaced with 
$\ln \Lambda_{SM}t$, allowing considerably larger values of $p$.
To summarize: there is a wide range of $p$ that gives a precise Higgs
mass prediction while allowing observers at $t_0$ to be common.

\section{Scanning the Top Quark Yukawa Coupling}

In the previous section we made an important assumption:  the only SM
parameters scanning in the landscape were those in the Higgs
potential.  
Of all the other SM parameters, the one in our patch of the universe that has the most
importance for electroweak symmetry breaking is the top quark Yukawa
coupling, $h$, via the RG equation for the scalar quartic coupling 
(\ref{eq:lambdarge}).  Hence we now discuss the consequences of
allowing $h$ to scan, leaving a discussion of allowing other Yukawa
couplings to scan until section 5.

Remarkably, we will show in this section that, allowing $h$ to scan, we can go
further than merely preserving the Higgs mass prediction of section 2;  using a 
landscape probability distribution, $P(\lambda_0,h_0)$, peaked in $\lambda_0$ 
beyond the SM metastability
boundary, we will be able to add a successful understanding of the top quark 
mass to our prediction for the Higgs mass.

In this section we again assume that the temperature was never above about 
$10^8$ GeV, at least for the patches of interest 
that dominate the probability distribution and evolved to the desired metastable 
phase,  so that we may ignore effects of vacuum transition by
thermal excitation.

\subsection{The Metastability Boundary in the $\lambda$--$h$ Plane}

When only $\lambda_0$ scans, the metastability boundary is given 
by the critical value of $\lambda_0$ that corresponds to 
$\lambda(M_{dom})=\lambda_c(M_{dom})$.
If $\lambda_0$ and $h_0$ both scan, 
what is the critical line for the metastability boundary in the 
$\lambda_0$--$h_0$ plane?  As noted in the introduction, it
has a special shape that will make an understanding of the top
quark mass possible.

\begin{figure}[!htbp]
\begin{center}
\begin{tabular}{l}
\includegraphics[width=.7\linewidth]{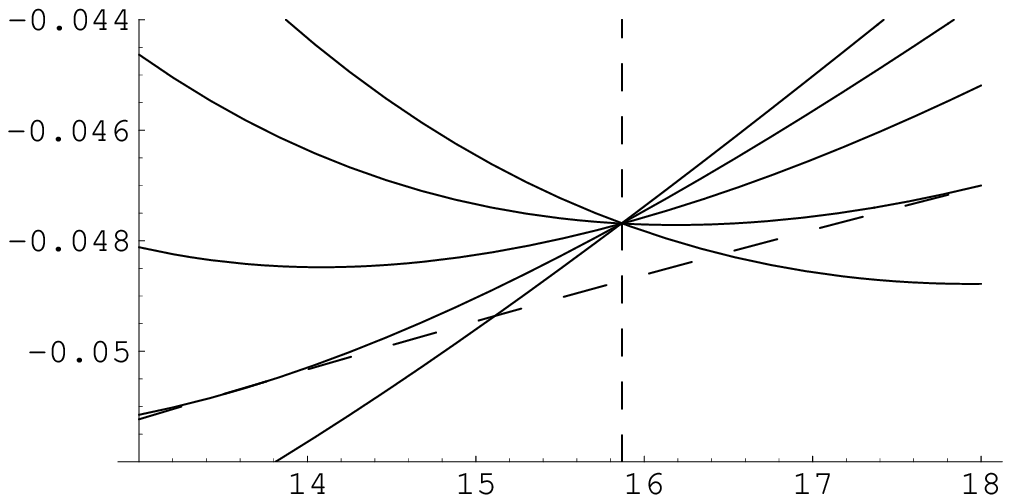} \vspace{1cm}
\\
\includegraphics[width=.7\linewidth]{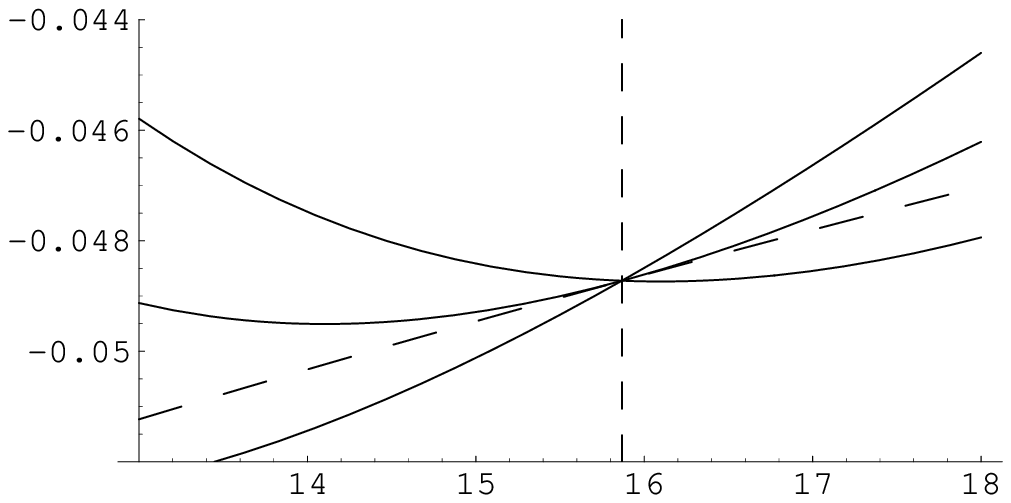} \vspace{1cm}
\end{tabular}
\begin{picture}(0,0)(400,200)
%
%
\Text(250,420)[b]{(a): $\quad \lambda(\Lambda) = -0.0477$}
\Text(250,210)[b]{(b): $\quad \lambda(\Lambda) = -0.0487$}
\Text(380,255)[lb]{$\log_{10} \mu$}
\Text(380,55)[lb]{$\log_{10} \mu$}
\Text(115,210)[b]{$\lambda$}
\Text(115,410)[b]{$\lambda$}
\Text(260,255)[lb]{$M_{dom}$}
\Text(260,55)[lb]{$M_{dom}$}
\Text(170,363)[b]{$A$}
\Text(155,340)[b]{$B$}
\Text(155,305)[br]{$C$}
\Text(155,270)[br]{$D$}
\Text(180,273)[tl]{$E$}
\Text(145,125)[br]{$b$}
\Text(140,100)[br]{$c$}
\Text(190,75)[tl]{$d$}
\end{picture}
\caption{\label{fig:RGdiffh} 
RG trajectories for $\lambda(\mu)$,
for different values of $h$ and a fixed value for $\lambda(\Lambda)$,
shown for a few decades in energy below $\Lambda_{SM} = 10^{18}$ GeV.
The scale $\Lambda$ is chosen to be $7.4 \times 10^{15}$ GeV,
corresponding to the value of $M_{dom}$ found in section 2 for the
special trajectory with $\lambda_0(\Lambda_{SM})=-0.0462$ that gives 
$m_t= 171$ GeV and $m_H$ = 106 GeV. 
The trajectories in the upper panel labeled $A,B,C,D$ and $E$ have $\lambda(\Lambda)=-0.0477$
with $h(\Lambda)=0.470, 0.447, 0.419, 0.391$ and $0.367$, respectively. 
Those in the lower panel, labeled $b, c$ and $d$,
are for $\lambda(\Lambda)=\lambda_c(\Lambda)=-0.0487$, with
$h(\Lambda)=0.447, 0.421$ and $0.391$.
In both panels the dashed line is the line of metastability.
The analysis uses SM RG equations at the 2-loop level. 
The gauge couplings at $\Lambda_{SM}$ are not scanned, but are fixed
to the values $(g_Y,g_L,g_s)(\Lambda_{SM})=(0.466,0.512,0.500)$
for all trajectories, the same as in Figure~\ref{fig:running4}. 
Uncertainties from higher loops and from the experimental uncertainty 
on $\alpha_s$ are not included.
}
\end{center}
\end{figure}

\begin{figure}
\begin{center}
\begin{tabular}{l}
\includegraphics[width=.85\linewidth]{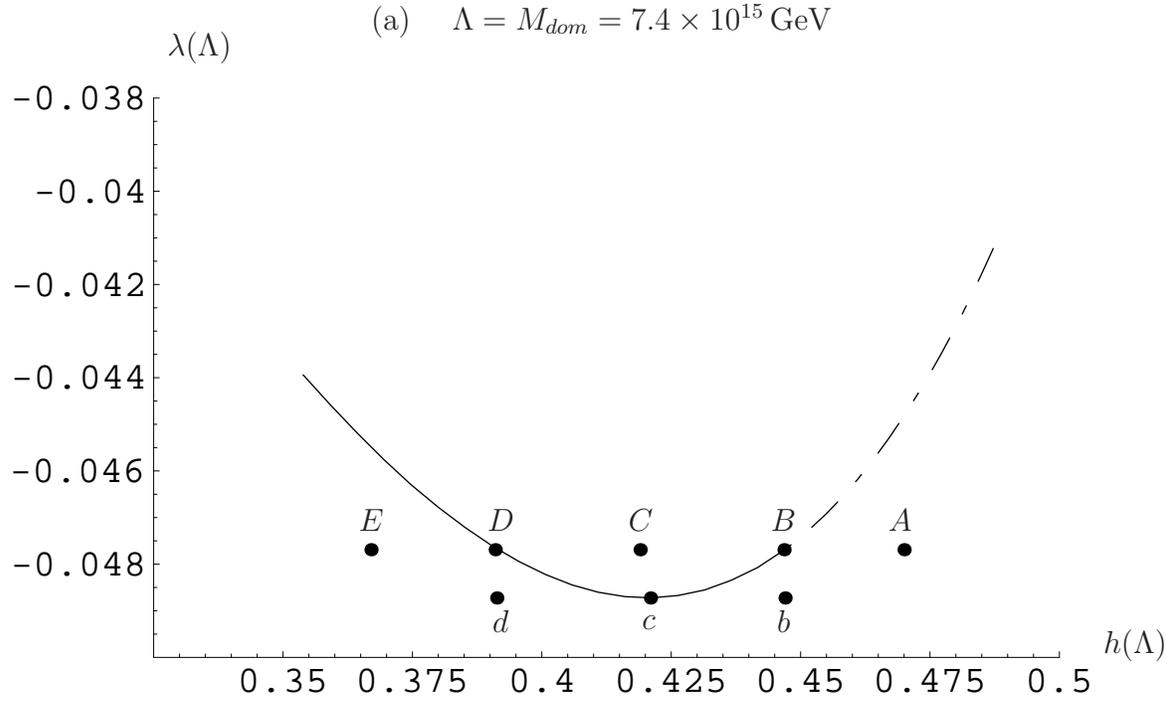}  \vspace{1.5cm} \\
\includegraphics[width=.85\linewidth]{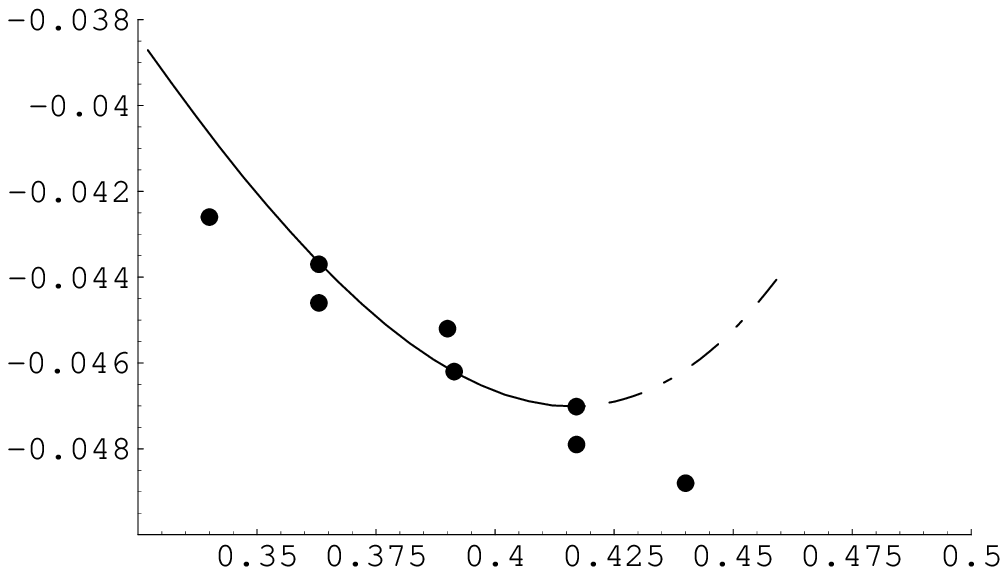}
\end{tabular}
\begin{picture}(0,0)(450,280)
%
%
\Text(250,570)[b]{(a) $\quad \Lambda = M_{dom}=7.4 \times 10^{15} \, \GEV$}
\Text(250,270)[b]{(b) $\quad \Lambda = \Lambda_{SM} = 10^{18}\, \GEV$}
\Text(98,560)[b]{$\lambda(\Lambda)$}
\Text(98,265)[b]{$\lambda(\Lambda)$}
\Text(440,340)[l]{$h(\Lambda)$}
\Text(440,45)[l]{$h(\Lambda)$}
%
\Text(163,383)[b]{$E$}
\Text(212,383)[b]{$D$}
\Text(265,383)[b]{$C$}
\Text(319,383)[b]{$B$}
\Text(363,383)[b]{$A$}
%
\Text(212,353)[t]{$d$}
\Text(268,351)[t]{$c$}
\Text(319,353)[t]{$b$}
%
\Text(111,172)[b]{$E$}
\Text(157,153)[b]{$D$}
\Text(210,127)[b]{$C$}
\Text(262,96)[b]{$B$}
\Text(305,65)[b]{$A$}
%
\Text(155,120)[t]{$d$}
\Text(210,93)[t]{$c$}
\Text(262,63)[t]{$b$}
\end{picture}
\caption{\label{fig:metastab3} The metastability boundary 
in the  $\lambda(\Lambda)$--$h(\Lambda)$ plane for (a) $\Lambda = M_{dom}$, found for  
the metastability trajectory of section 2, and 
(b) $\Lambda = \Lambda_{SM} = 10^{18}$ GeV.  
The points labeled $A,B,C,D,E$ and $b,c,d$ correspond to the trajectories 
of Figure \ref{fig:RGdiffh}. The metastability boundary is shown dashed in regions 
where the calculation is unreliable; using SM equations, the dominant scale for 
tunneling is found to be above $\Lambda_{SM}= 10^{18}$ GeV. }
\end{center}
\end{figure}

Consider a scale $\Lambda \leq \Lambda_{SM}$, and fix a value of $\lambda(\Lambda)$.
Figure~\ref{fig:RGdiffh}a 
shows trajectories of $\lambda$, starting from this initial condition, for 
various values of $h$. In the figure,  $\Lambda = M_{dom}$ was used, but the qualitative 
picture is independent of this choice.
When the top Yukawa is too small or too large, $\lambda$
runs into the forbidden unstable region.   As $\lambda(\Lambda)$ becomes smaller, 
the allowed range of $h$ shrinks accordingly. When $\lambda(\Lambda)$ actually lies on
the metastability boundary itself, as in Figure~\ref{fig:RGdiffh}b, 
the only permissible value of $h$ remaining is the
one for which $\lambda$ runs parallel to the boundary at the chosen 
scale, trajectory $c$ of Figure~\ref{fig:RGdiffh}b.  The resulting 
metastability boundary in the $\lambda(\Lambda)$--$h(\Lambda)$ plane is shown in 
Figure \ref{fig:metastab3}a.  The trajectories
labeled $A,B,C,D,E$ and $b,c,d$ in Figure \ref{fig:RGdiffh} appear as points in Figure
\ref{fig:metastab3}. The special trajectory $c$ in  Figure
\ref{fig:RGdiffh}b is clearly at a special location on the
metastability boundary in Figure \ref{fig:metastab3}a.
If there is a probability pressure 
in the landscape pushing $\lambda(\Lambda)$ 
towards $\lambda_c(\Lambda)$, as in section 2, then we may simultaneously obtain the prediction
that $h(\Lambda)$ will lie at the critical value corresponding to the
trajectory $c$.

Now, the low energy value of the top mass corresponding to the critical point depends on the 
actual scale $\Lambda$ we choose.  If $\Lambda$ is increased, the general shape 
of the metastability boundary remains the same, as shown in Figure
\ref{fig:metastab3}b for $\Lambda = 10^{18}$ GeV, but the special
trajectory has changed from $c$ to near $B$, with a consequent change
in the top mass prediction.
The prediction for $h$ should be made at the scale at which there is peaking 
in the probability distribution pushing towards the critical point. 
Although with $h$ scanning, this is not a RG invariant 
requirement, it is a reasonable possibility for this to occur close to the 
scale of new physics, in particular, at the SM cutoff $\Lambda_{SM}$.
What is remarkable is that if we take $\Lambda = \Lambda_{SM}$ to be within a 
few orders of magnitude of the Planck scale, then the corresponding prediction 
for $h$ yields a value for the the top mass that agrees closely with experiment.  
The uncertainty from varying  $\Lambda = \Lambda_{SM}$ is mild.

The astute reader will notice a problem with this analysis; the $\lambda- h$ metastability
boundary was derived using the SM RG equations and tunneling rate calculations, which are 
expected to break down above the scale $\Lambda_{SM}$.  
As a result of effects from the new physics, the $\lambda_0$--$h_0$ metastability
boundary will differ from the form shown in Figure \ref{fig:metastab3}, and is sketched in 
Figure \ref{fig:metastab4}.  
However, the possible deviations are greatly restricted: only the half of the curve 
to the right of the critical point, corresponding to tunneling events to scales 
beyond $\Lambda_{SM}$, will be affected.  Moreover, this part of the curve cannot 
change in an arbitrary way;  $\lambda_{0,c}$ is still a strict lower bound on the 
Higgs quartic at $\Lambda_{SM}$, and thus the true curve must lie
above the dashed line of Figure \ref{fig:metastab4}.  As long as the true curve 
rises somewhat above this horizontal line, the central value of our 
prediction will be unchanged.  If this curve did remain too
flat, however, our prediction would have a large error 
bar on the high side; $m_t$ would have a lower bound, but not much of an upper 
bound.  This situation could arise if tunneling events
to scales larger than $\Lambda_{SM}$ are very suppressed.\footnote{The 
tunneling rate depends exponentially on the action.  The reason that the 
metastability boundary changes so gradually in the SM is because of tree-level 
scale independence.  However, the tunneling rate of the full theory is likely 
to differ substantially from that of the SM, so that we expect the actual behavior
to be more extreme than the two dashed curves of Figure \ref{fig:metastab4}.}  
To obtain a well controlled prediction for $m_t$,
we will thus make one additional assumption in this section: either
\begin{enumerate}
 \item The metastability boundary rises sufficiently rapidly above $h_{0,c}$ 
so as to provide a reasonably sharp upper bound on the top mass, or
 \item There is a fundamental probability distribution pressure in the landscape 
pushing $h_0$ towards smaller values, and hence towards $h_{0,c}$.  We take $q =  
- \frac{\partial \ln P(\lambda_0,h_0)}{\partial \ln h_0}|_c$, and require 
$1 \simlt q \leq p$.
\end{enumerate}

\subsection{The Critical Value of the Top Mass}

\begin{figure}[!htbp]
\begin{center}
\includegraphics[width=0.9\linewidth]{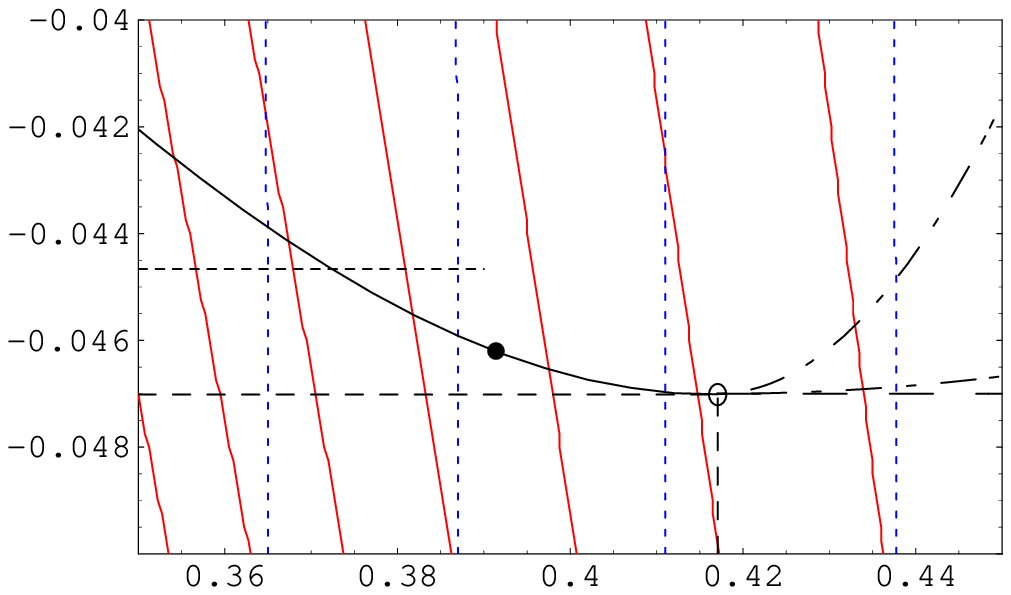}
\begin{picture}(0,0)(250,0)
%
%
\Text(250,25)[l]{$h_0$}
\Text(-130,265)[b]{$\lambda_0$}
\Text(265,90)[l]{$\lambda_{0,c}$}
\LongArrow(260,92)(250,92)
\Text(123,0)[t]{$h_{0,c}$}
\LongArrow(123,5)(123,20)
%
\Text(-125,45)[l]{\color{red} 70}
\Text(-90,45)[l]{\color{red} 80} 
\Text(-50,45)[l]{\color{red} 90}
\Text(-5,45)[l]{\color{red} 100}
\Text(50,45)[l]{\color{red} 110}
\Text(110,45)[l]{\color{red} 120}
\Text(180,45)[l]{\color{red} 130}
\Text(-70,240)[l]{\color{blue} 165}
\Text(10,240)[l]{\color{blue} 170}
\Text(105,240)[l]{\color{blue} 175}
\Text(200,240)[l]{\color{blue} 180}
%
\LongArrow(-30,95)(-30,145)
\Text(-27,130)[l]{$\frac{|\lambda_{0,c}|}{p}$}
\end{picture}
\caption{\label{fig:metastab4} The metastability boundary 
in the $\lambda_0$--$h_0$ plane for $\Lambda_{SM} = 10^{18}$ GeV. 
For $h_0 < h_{0,c}$ the solid curve is reliably 
computed in the SM. For $h_0 > h_{0,c}$, the metastability boundary must lie 
above the dashed horizontal line.  Two possible boundaries are shown; the lower 
(upper) dot-dash curve results when tunneling in the full theory is suppressed 
(rapid). The nearly vertical blue dotted lines are contours of the top mass 
in 2.5 GeV intervals, while the red solid lines, also not far from vertical, 
are contours of the Higgs mass with 10 GeV intervals, 
shown for $\alpha_s = 0.1176$.  The open circle is the 
critical point, giving the most probable observed values of $m_t$ and $m_H$  
in the multiverse, corresponding to trajectory $B$ of Figure \ref{fig:metastab3}b.
The closed circle represents a patch of the multiverse on the metastability 
boundary where the top quark is measured to be 171 GeV. In our patch, the top
mass is within a couple of GeV of this number, and hence lies somewhere between 
the $m_t$ of 170 GeV and 175 GeV contours.  The importance for the Higgs mass
prediction of decreasing the experimental uncertainties on $m_t$ is clear. 
An arrow denotes the 
magnitude of the expected upward fluctuations of the quartic coupling 
$\lambda_0$ above its critical value for peaking parameter $p=20$.   
Such a fluctuation above $\lambda_{0,c}$ would lead to a point on the metastability
boundary where the top mass is about 171 GeV and the Higgs mass about 110 GeV.  
}
\end{center}
\end{figure}

The critical value of the quartic coupling at any scale $\mu$,
$\lambda_c(\mu)$, is defined by $\Gamma(\mu) \tau^4 = 1$, 
where $\tau = 10^{10}$ years. 
By taking a derivative with respect to $\mu$ and using (\ref{eq:rateM}), 
we obtain
%
 \begin{equation}
 16\pi^2 \frac{d\lambda_c}{d\ln \mu}= 24 \lambda_c^2.
 \end{equation}
On the other hand, the critical RG trajectory for $\lambda$, with h scanning, touches the
curve $\lambda_c(\mu)$ tangentially at $\Lambda_{SM}$.  In this way, the condition for
the critical top Yukawa coupling $h_c$ follows from
 \begin{equation}
\left. \frac{d \lambda}{d\ln \mu} \right|_{\lambda_0 = \lambda_{0,c}, \mu = \Lambda_{SM}} = 
\left. \frac{d \lambda_c}{d\ln \mu} \right|_{\mu = \Lambda_{SM}},
 \end{equation}
giving
\begin{equation}
\left( 12 \lambda_c h_c^2 - 6 h_c^4 - 9 \lambda_c g_L^2 - 3 \lambda_c g_Y^2 
  + \frac{3}{8} g_Y^4 + \frac{3}{4} g_L^2 g_Y^2 + \frac{9}{8} g_L^4 
     \right)_{\Lambda_{SM}} =0.
\label{eq:hc1}
\end{equation}
%
%
%
Since $\lambda_{c,0} < 0$, the first 2 terms in (\ref{eq:hc1}) are negative
while all the rest are positive. Taking $\lambda_{c,0}$
from equation (\ref{eq:lambdac}), and using two loop SM RG running for 
$g_L(\Lambda_{SM})$ and $g_Y(\Lambda_{SM})$,\footnote{The choice of the 
Higgs boson mass as a boundary condition of the RG equation 
has negligible impact on $g_L(\Lambda_{SM})$ and $g_Y(\Lambda_{SM})$.} 
we find the critical value for
the top coupling $h_{c,0} = h_c(\Lambda_{SM})$.  For example, for $\Lambda_{SM} =
10^{18}$ GeV, $h_{c,0} = 0.417$.
%
RG scaling to the weak scale, we find a critical top quark mass
%
\begin{equation}
m_{t_c} = \left( 176.2 \pm 3 + 2.2 \log_{10} \frac{\Lambda_{SM}}{10^{18} \GEV} \right) \GEV.
\label{eq:mt-crit}
\end{equation}
where the uncertainty $\pm 3 \, \GEV$ comes from higher order contributions to RG 
scaling and to threshold corrections at the scale of the top quark mass.
This critical point is denoted by an open circle in Figure \ref{fig:metastab4}. 
Contours of $m_t$ and $m_H$ are also shown in this figure by the near vertical
blue dotted and red solid lines. Clearly a crucial question is how far from
the critical point a typical patch in the false vacuum is likely to be.

\subsection{The Width of the Top Mass Prediction}

We would next like to determine how far from $m_{t_c}$ the top mass 
could go, $\Delta m_{t \pm}$, before 
the probability would have fallen by a factor of $e$.  In order to lower the top Yukawa $h_0$
below its critical value $h_{0,c}$, the Higgs quartic must be raised to prevent 
the RG flow of $\lambda$ from heading into the forbidden region.  
This costs probability through the Higgs quartic 
probability distribution. As in section 2, we lose a factor of $e$ 
in probability when the Higgs quartic is raised above the metastability 
boundary by an amount ${|\lambda_c|}/p$.  The lower bound on the top 
mass prediction then corresponds to the smallest top Yukawa coupling 
compatible with this raised value of $\lambda$, and can be read off from the solid curve
of Figure \ref{fig:metastab4}.
For a small deviation from the critical point $(\lambda_{0,c},h_{0,c}) =
(-0.047,0.417)$, the metastability boundary rises as
$(\lambda_0-\lambda_{0,c}) \simeq 1.2 (h_0 - h_{0,c})^2$, and the
downward shift $\Delta h_0 \sim \sqrt{|\lambda_{0,c}|/(1.2 p)}$ or 
$\Delta h_{0,c}/h_{0,c} \sim 0.47/\sqrt{p}$
gives the lower bound.
Running down to the weak scale, $\Delta h/h$ shrinks by a factor of 0.4,
and we find a corresponding downward shift in the top mass of 
\begin{equation}
\Delta m_{t-} \approx \frac{35 \, \GEV}{\sqrt{p}}.
\label{eq:mt-}
\end{equation}
Such a shift is illustrated and discussed in Figure \ref{fig:metastab4}
for $p=20$.

The upper bound on the top mass prediction is less certain; it depends on either the shape of the $\lambda-h$ metastability
boundary to the right of the critical point, or on the strength, given by $q$, of the the fundamental probability distribution
on $h$.  
In the former case, the upper bound depends on unknown physics beyond 
$\Lambda_{SM}$ and we cannot make a quantitative statement. In the latter case
the probability will have dropped by a factor of $e$ for a top Yukawa 
that is above the critical value by an amount $\approx h_c/q$,
corresponding to
\begin{equation}
\Delta m_{t+} \approx \frac{75 \, \GEV}{q}.
\label{eq:mt+}
\end{equation}
Putting everything together, our final result for the top mass prediction is 
\begin{equation}
m_t = \left( 176.2 \pm 3 + 2.2 \log_{10} \frac{\Lambda_{SM}}{10^{18} \GEV} 
-  \frac{3}{\sqrt{p/135}}   + \frac{3}{q/25} \right) \GEV.
\label{eq:mt}
\end{equation}
We find this to be a striking success. Indeed, for the agreement with
experiment not to be accidental, the a priori
distribution for $\lambda_0$ must be highly peaked, 
certainly with $p \simgt 20$, so that the uncertainty in the 
Higgs mass prediction from the landscape, 
(\ref{eq:DeltamH}), is less than 1 GeV. 

\section{Higgs Mass Prediction from Thermal Fluctuations}

All the analysis in the preceding sections assumed that the temperature 
of the thermal plasma remained sufficiently small after the last era
of inflation. On the other hand,  thermal fluctuations can themselves
induce vacuum transitions from the false vacuum 
with $\vev{H}=v$ to the true vacuum with large $\vev{H}$. If the 
temperature was sufficiently high, then the vacuum transitions are dominated 
by these thermal processes, not by the quantum tunneling discussed so
far \cite{Anderson, AV, Sher, EQ}.
This section investigates how the results in the earlier sections should be 
modified in this case.  We will continue to assume that the Standard Model is a good 
approximation up to a very high scale, and will return to the assumption that the top quark Yukawa
 coupling does not effectively scan in the landscape.

\subsection{ Vacuum Transitions at High Temperatures}

As the inflaton decays, the inflation vacuum energy density $V_I$ is converted into a thermal plasma.  The maximum temperature
of this plasma will be denoted by $T_{\rm max}$, which we will assume to be less than $\Lambda_{SM}$.   
The reheating temperature, at which the thermal plasma dominates the energy density of the 
universe, will be denoted by $T_R$.  In the simplest scenarios for reheating, we have the relation $T_{\rm max}^4 \sim \sqrt{T_R^4 V_I}$.
Note that since tensor perturbations have not yet been observed in the cosmic microwave background, an upper bound has been placed on $V_I$
 of about $(10^{16} \, \GEV)^4$.

The vacuum decay rate induced by thermal fluctuations is again calculated 
by expanding around a bounce solution. The Higgs potential now
involves a 
thermal contribution in addition to the one at $T=0$:
\begin{equation}
V_{\rm tot}(H) = \frac{\lambda(H)}{4} H^4 + V_{TH}(H;T).
\label{eq:th-pot}
\end{equation}
The quartic potential with negative $\lambda$ does not have an
SO(3)-symmetric bounce solution. A bounce exists only for the total 
potential (\ref{eq:th-pot}). The potential does not have a conformal 
symmetry at all, and as a result, 
bounces to field values of order $T$ dominate in the decay rate, and the bounce action depends only on $\lambda(T)$.
This is to be contrasted with the situation in section 2, in which tunneling to all scales had to be 
summed up, and the entire RG trajectory $\lambda(\mu)$ was important for tunneling at any given time.  It should be noted,
though, that since the temperature of the universe changes, the full RG
behavior of $\lambda(\mu)$ up to $\mu = T_{\rm max}$ will still be important
in this section, through the time dependence of the temperature.

We will now perform an analysis at a somewhat low level of approximation to get 
an overall qualitative picture of what is going on.  For precise quantitative
 results, numerical calculations will be used, as can be seen in what follows.
We will employ the approximation adopted by \cite{AV}; 
for $g_{\rm eff} H \ll T$, the high-temperature ($H/T$) expansion 
of the potential may be used.  The leading
term of the 1-loop thermal potential is then the  thermal mass term
\begin{equation}
V^{(2)}_{\rm TH} \simeq \frac{1}{2}g^2_{\rm eff} T^2 H^2,
\label{eq:thmass}
\end{equation}
where 
\begin{equation}
g_{\rm eff}^2 \equiv \frac{1}{12}\left( 
  \frac{3}{4}g_Y^2 + \frac{9}{4}g_L^2 + 3 h^2 \right).
\label{eq:geffdef}
\end{equation}
For $\lambda(T)$ negative and  
$g_{\rm eff}^4 \ll |\lambda(T)|$, the potential barrier for the transition 
occurs at around $H \sim g_{\rm eff} T/\sqrt{|\lambda(T)|}$.  Thus $g_{\rm eff} H$ is in this case
much less than $T$, so that the high-temperature expansion is a good 
approximation. Reference \cite{AV} then obtained  
\begin{equation}
\frac{S_3}{T} \simeq \frac{6.015 \pi g_{\rm eff}(T)}{|\lambda(T)|},
\label{eq:SoverT}
\end{equation}
and a corresponding thermal decay rate per unit volume of 
\begin{equation}
\Gamma[\lambda(T); T] \simeq T^4 \left(\frac{S_3}{2\pi T} \right)^{3/2} 
e^{-\frac{S_3}{T}}.
\label{eq:rateT}
\end{equation}

In order to determine the probability of a point remaining in the false vacuum
at a time $t$, one must consider the possibility of bubble nucleations throughout
the point's past light-cone.  The result for the fraction of the universe in the false vacuum
at time $t$ is then \cite{GW}
\begin{equation}
f(\lambda_0; t) = \exp \left(- \int^t_{t_{\rm init}} dt_1 \, 
  \Gamma[\lambda(T(t_1)); T(t_1)] a(t_1)^3 \frac{4\pi}{3} 
    \left\{ \int^t_{t_1} \frac{d t_2}{a(t_2)} \right\}^3 
                         \right),
\end{equation}
where $t_{\rm init}$ is the time right after the end of inflation, and 
$a(t)$ is the scale factor of the expansion of the universe.
In fact, the same equation was used in section 2 as a precursor to (\ref{eq:GW}).  Here, however, 
 $\Gamma$ changes as the temperature 
of the universe changes, and so we must now use the more general expression. Assuming 
a standard thermal history after the end of inflation, 
namely, a matter-dominated era of inflaton coherent oscillations, 
followed by radiation domination and matter domination (and further 
dark-energy domination), we obtain 
\begin{eqnarray}
f(\lambda_0; t) & \sim & \exp \left(- \int^{T_{\rm max}}_{T_R} \frac{dT}{T}
 \Gamma[\lambda(T); T]
 \frac{M_{Pl}T_R^2}{T^4}\left(\frac{T_R}{T}\right)^{8}
 \left(\frac{T(t)}{T_R}\right)^3 t^3 \right. \nonumber \\
& & \qquad \left. \; \; -\int^{T_R}_{\lambda(T) < 0} \; \; \frac{dT}{T}
 \Gamma[\lambda(T); T] \; \;
 \frac{M_{Pl}}{T^2} \qquad \quad 
 \left(\frac{T(t)}{T}\right)^3 t^3 
\right), 
\label{eq:fFthermal}
\end{eqnarray}
assuming that $t$ is in the recent matter-domination era.\footnote{If $t$ is 
in the era of cosmological constant dominance that follows after the recent 
matter-dominance era, then $T(t)^3 t^3$ is replaced by 
$T^{'3}_{eq} H_0^{-3}$ for $t \gg t'_{eq}$, 
where $T'_{eq}$  and $t'_{eq}$ are the temperature and epoch of 
matter-cosmological constant equality, and $H_0$ is the Hubble constant 
of the cosmological constant. Note that the $t$-dependence completely 
disappears from $f(\lambda_0; t)$.} 
Here,  $T_{eq}$ is the temperature of matter-radiation equality, and 
$T(t)$ is the photon temperature at time $t$.
We have used $T^4 \propto 1/a(t)^{3/2}$ and 
$dt \sim - (M_{Pl} T_R^2/T^4)(dT/T)$ during the inflaton dominated era \cite{KT}.
Coefficients of order unity are ignored in the above expression for $f$, 
for the same reason as in section 2.

\begin{figure}
\begin{center}
\begin{tabular}{c}
\includegraphics[width=.75\linewidth]{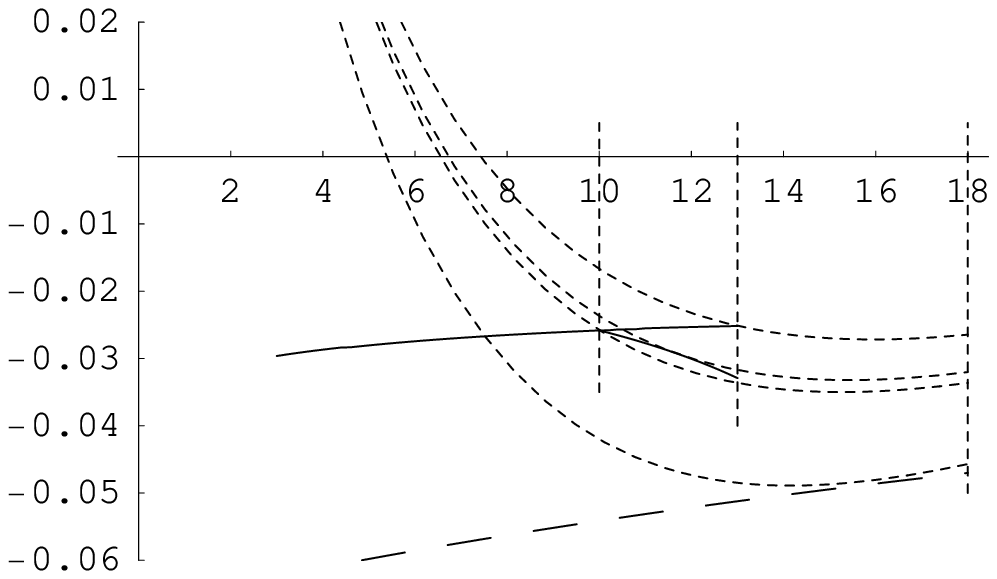}  \vspace{1.5cm}
\\ 
\includegraphics[width=.6\linewidth]{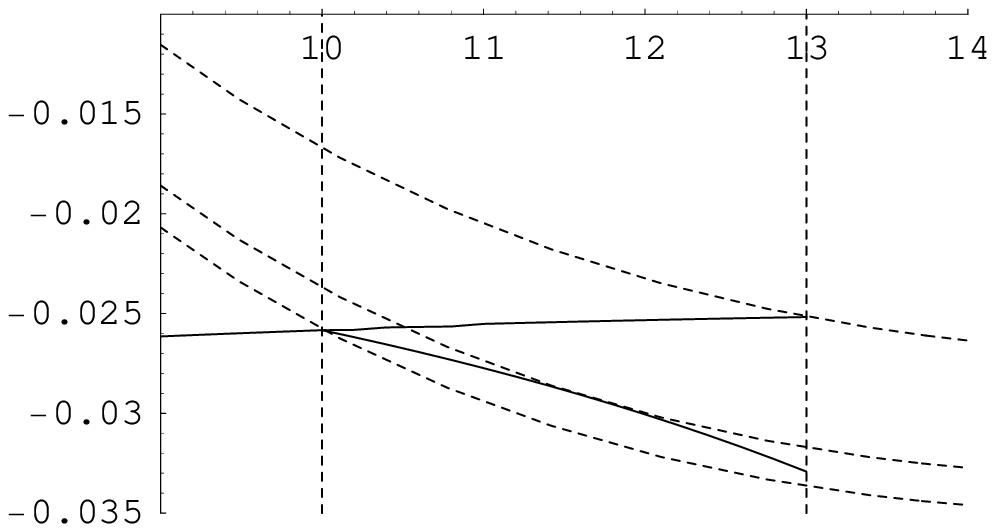} 
\end{tabular}
\begin{picture}(0,0)(400, 230)
%
%
%
\Text(0,450)[lt]{$(a)$}
\Text(395,395)[l]{$\log_{10} \mu$}
\Text(75,460)[b]{$\lambda$}
\Text(25,200)[lt]{$(b)$}
\Text(360,180)[l]{$\log_{10} \mu$, $\log_{10} T$}
\Text(110,190)[b]{$\lambda$}
\Text(245,415)[b]{$T_R^{(L)}$}
\Text(285,420)[l]{$T_{\rm max}$}
\Text(285,438)[l]{$T_R^{(I)},$}
\Text(373,415)[b]{$\Lambda_{SM}$}
\Text(150,328)[b]{$\lambda_c(T)$}
\Text(175,259)[b]{$\lambda_c(M)$}
\Text(250,285)[rt]{\color{red} 106.3}
\Text(320,309)[t]{\color{red} 113.1}
\Text(335,316)[b]{\color{red} 113.9}
\Text(350,333)[b]{\color{red} 116.5}
\Text(158,180)[b]{$T_R^{(L)}$}
\Text(295,180)[b]{$T_R^{(I)}$, $T_{\rm max}$}
\Text(260,59)[rt]{$\lambda_c^{(L)}(T)$}
\Text(280,83)[rt]{$\lambda_c^{(I)}(T)$}
\Text(350,80)[l]{\color{red} $m_H = 116.5$ GeV}
\Text(350,43)[l]{\color{red} $m_H = 113.9$ GeV}
\Text(350,33)[l]{\color{red} $m_H = 113.1$ GeV}
\end{picture}
\caption{\label{fig:criticalT} The critical line of stability due to
vacuum decay by thermal fluctuations, shown as solid lines terminating
at $T_{\rm max}$. Two cases are shown, both with $T_{\rm max} = 10^{13}$ GeV.
The upper solid line, case (I), corresponds to instantaneous reheating with
$T_R = T_{\rm max}$.  Varying $T_R = T_{\rm max}$ away from $10^{13}$ GeV
will simply change the termination point of the line.    
The lower solid line, case (L), corresponds to late 
reheating with $T_R = 10^{10}$ GeV. Varying $T_R$ will alter the position 
of the bend in the line, while altering $T_{\rm max}$ will alter the 
termination point.
The two lines coincide for $T < 10^{10}$ GeV.  
The long dashed line near the bottom is $\lambda_c(M)$ discussed 
in section 2. 
Four RG trajectories of $\lambda(\mu)$ are also shown, as dotted lines, 
with Higgs masses of (106.3, 113.1, 113.9, 116.5) GeV.
The lower panel is a blow-up of the upper panel.
The low temperature value of the critical Higgs mass, 106.3 GeV,
is raised to 113.9 GeV in case (L) and 116.5 GeV in case (I).
}
\end{center}
\end{figure}

One can introduce a critical value of the Higgs quartic coupling 
$\lambda_c(T)$ by 
\begin{eqnarray}
\Gamma[\lambda_c(T); T] \tau^3   \frac{M_{Pl}}{T^2} 
 \left(\frac{T_R}{T}\right)^{10}\left(\frac{T_0}{T_R}\right)^3 \simeq 1, & & 
T_R < T, \\
\Gamma[\lambda_c(T); T] \tau^3 \frac{M_{Pl}}{T^2} \qquad \qquad 
\left(\frac{T_0}{T}\right)^3 \simeq 1, & & T < T_R,
\end{eqnarray}
where\footnote{Factors of order unity are not important here, since 
these equations balance a very small $\Gamma$ against a 
very large $\tau^3 M_{Pl}/T^2$. In particular, it is not important 
whether we are in the dark-energy dominated era or not, or which values 
to use for $\tau$ and $T_0$. A factor of 2--3 changes $\lambda_c(T)$ by 
half percent, which corresponds to less than $0.1$ GeV change in the Higgs 
boson mass.}
$T_0 \simeq 2.73 \, {}^{o} {\rm K} \simeq 2.35 \times 10^{-4} \, \EV$
and $\tau = 10^{10}$ yrs.
By taking a logarithm and using (\ref{eq:rateT}), we have\footnote{
\label{fn:det} The number 
``243'' contains $(3/2) \ln (243/2\pi) \simeq 5.5$ that comes from 
the 1-loop functional determinant. Note that $\Delta S$ of section 
2, also calculated from a 1-loop functional determinant, contributed 
a similar order of magnitude, $\sim -10$ \cite{Strumia}.}
\begin{eqnarray}
 \frac{6.0 \pi g_{\rm eff}(T)} { |\lambda_c(T)| } & \simeq &
 \ln \left( \frac{ M_{Pl} }{T} (\tau T_0)^3  \right)
 + \frac{3}{2}\ln \left(\frac{S_3}{2\pi T} \right) 
+ \left[ 7 \ln \left(\frac{T_R}{T} \right) \right], \\
\mbox{or equivalently} \hspace{1cm} \lambda_c(T) & \simeq & -
 \frac{ 6.0 \pi g_{\rm eff}(T)}{243 - \ln \frac{T}{v} 
        - \left[7 \ln \frac{T}{T_R} \right]}.
\label{eq:lambdacT}
\end{eqnarray}
The term $[7 \ln (T/T_R)]$ is in square brackets because it should
be included only when $T > T_R$.
The critical line of stability $\lambda_c(T)$ is shown in Figure 
\ref{fig:criticalT}, along with some RG trajectories of the Higgs quartic 
coupling for $m_t=171$ GeV.
If an RG trajectory passes below the critical line at an energy scale 
$\mu < T_{\rm max}$, we will have $f(\lambda_0;t) \ll 1$, so that 
most patches of the universe with the corresponding Higgs boson mass would have decayed to the 
true vacuum by now. 

In Eq. (\ref{eq:lambdacT}), $g_{\rm eff}(T)$, as calculated from (\ref{eq:geffdef}), lies in
the range $0.3$--$0.4$ for a wide range of $T$.  At the same time,  
$|\lambda_c(T)|$ ranges from $0.03$--$0.04$, so that $|\lambda_c(T)|$ is not much 
bigger than $g_{\rm eff}^4$.  For this reason, the high-temperature expansion of the
thermal potential is not particularly good, and  consequently, 
$6.0 \pi g_{\rm eff}/|\lambda|$
is not actually a very good estimate for the bounce action $S_3/T$ \cite{EQ}. The critical 
line of stability in Figure \ref{fig:criticalT} was therefore obtained from a numerical 
calculation, the details of which are explained in the appendix.
On the other hand, expression (\ref{eq:lambdacT}) does capture 
the qualitative behavior of $\lambda_c(T)$ shown in the figure. For $T < T_R$, 
the asymptotically free behavior of the SU(2)$_L$ and top Yukawa couplings
 are more important than the $-(1/243) \ln (T/v)$ term in the denominator, 
and $|\lambda_c(T)|$ slowly decreases as $T$ increases. For $T > T_R$, 
however, the extra $-(7/243) \ln (T/T_R)$ term in the denominator is more important, 
and $|\lambda_c(T)|$ increases. 

Figure \ref{fig:criticalT} shows $\lambda_c(T)$ for two scenarios: 
one for $T_{\rm max} = T_R = 10^{13} \, \GEV$, and the other for 
$T_{R} = 10^{10} \, \GEV$ and $T_{\rm max} = 10^{13} \, \GEV$.
In the instantaneous reheating scenario at $T=10^{13}$ GeV (I), an RG trajectory 
of $\lambda(\mu)$ with $m_H=116.5$ GeV (upper dotted curve in 
Figure \ref{fig:criticalT}b) touches the line of $\lambda^{(I)}_c(T)$ 
at $T \simeq 10^{13}$ GeV.   Patches of the universe with Higgs 
boson masses smaller than $116.5 \, \GEV$ would have mostly decayed because their 
RG trajectories pass below the $\lambda^{(I)}_c(T)$ line 
at some temperature less than $T_{\rm max}$.
In the late reheating scenario with $T_R = 10^{10}$ GeV (L), the RG trajectory 
for $m_H = 113.9$ GeV (middle dotted curve in Figure \ref{fig:criticalT}b) 
touches $\lambda^{(L)}_c(T)$ at a temperature in between $T_R$ and $T_{\max}$. 
Thus, in this case patches of the universe with $m_H<113.9$ GeV typically
 decayed to the true vacuum within the inflaton dominated era. 
It is clear from Figure \ref{fig:criticalT}a that the constraint from 
 vacuum decay due to quantum fluctuations, shown as a long dashed line, 
is of negligible importance for both (I) and (L) scenarios discussed above,
as well as many other cases with high $T_R$.

\subsection{The High Temperature Prediction for the Higgs Boson Mass}

When the a priori distribution $P(\lambda_0)$ of the Higgs quartic coupling 
 is highly peaked toward negative values, a balance between 
$f(\lambda_0;t)$ and $P(\lambda_0)$ determines the most likely value 
of the Higgs boson mass. Since the double-exponential cut-off in $f$ 
is so sharp, again the point of maximum probability occurs essentially at the cut-off, 
as in section 2.  Thus, the phenomenological lower bound on the Higgs boson mass may again  become 
the prediction of a landscape scenario.

\begin{figure}[ht]
\begin{center}
\includegraphics[width=.7\linewidth]{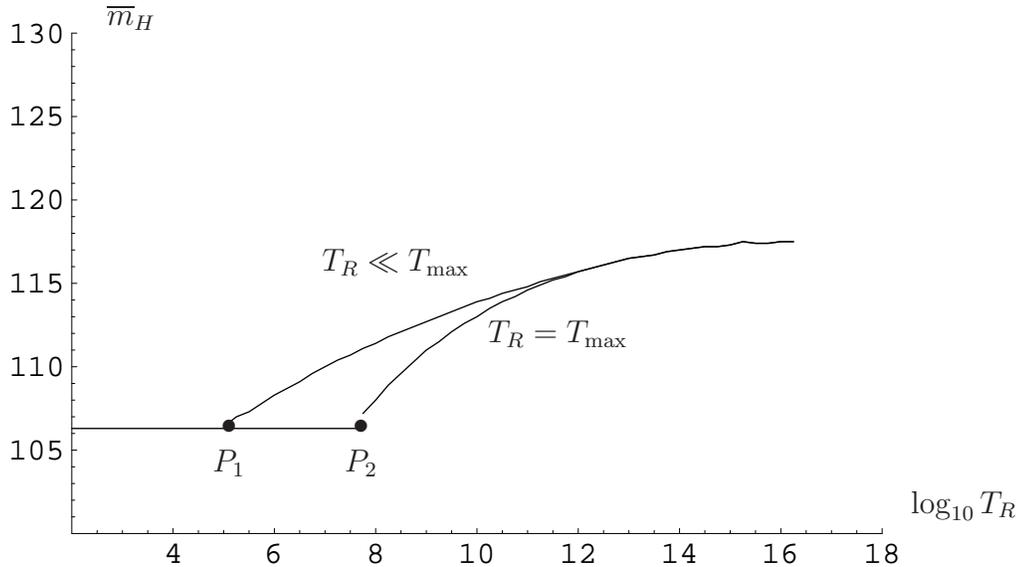} 
\begin{picture}(0,0)(400,0)
%
%
%
\Text(400,25)[l]{$\log_{10} T_R$}
\Text(105,205)[b]{$\overline{m}_H$}
\Text(233,112)[rb]{$T_R \ll T_{\rm max}$}
\Text(240,95)[lt]{$T_R = T_{\rm max}$}
%
\Text(142,55)[c]{$\bullet$}
\Text(142,46)[t]{$P_1$}
%
\Text(192,55)[c]{$\bullet$}
\Text(192,46)[t]{$P_2$}
\end{picture}
\caption{\label{fig:mhprediction} The prediction for the central value
  of the Higgs 
boson mass as a function of the reheating temperature $T_R$. The upper line 
is for the $T_{\rm max} \gg T_R$ scenario, and the lower line 
for the instantaneous reheating scenario, $T_{\rm max} = T_R$. 
We assume that the top Yukawa coupling is fixed in the landscape, so that 
$m_t = 171$ GeV. Note that the $\pm 6$ GeV error bar in (\ref{eq:pred-val}), 
coming from threshold corrections at the weak scale etc., is not shown.}
\end{center}
\end{figure}

Let us now consider a case with $T_R = 10^{10} \, \GEV$ as an example. 
If $T_{\rm max}$ is also $10^{10} \, \GEV$, 
then $f$ sets a cut-off $\lambda(10^{10} \, \GEV) > \lambda_c(10^{10} \, \GEV) 
\simeq -0.026$, as we see from Figure \ref{fig:criticalT}b. The 
RG trajectory with $\lambda(\mu = 10^{10} \, \GEV) = -0.026$ becomes the 
landscape prediction, which corresponds to $\overline{m}_H = 113.1$ GeV. 
If $T_{\rm max}$ is a little higher than $10^{10}$ GeV, patches 
with $m_H = 113.1$ GeV have mostly decayed during the inflaton-oscillation 
dominated era, because the RG trajectory lies below the critical line 
of stability $\lambda_c^{(L)}(T)$. Thus the  prediction for $m_H$ goes up as $T_{\rm max}$ 
increases. For sufficiently high $T_{\rm max}$, however, this sensitivity to $T_{\rm max}$
disappears; the RG trajectory with $m_H = 113.9$ GeV 
touches $\lambda^{(L)}_c(T)$  at around $10^{12} \, \GEV$.
As long as $T_{\rm max}$ is higher than this temperature, the landscape 
prediction will remain at $\overline{m}_H = 113.9 \, \GEV$. 
Figure \ref{fig:mhprediction} shows the prediction for the most 
probable Higgs mass $\overline{m}_H$ in these 
 two limits, the instantaneous reheating scenario, with $T_{\rm max} = T_R$, and the high  
$T_{\rm max} \gg T_R$ scenario. The prediction will always lie 
in between these two lines for any $T_{\rm max}$.
 
Now, for sufficiently high reheating temperature, $T_R\simgt 10^{13}$ GeV, the prediction does 
not depend on $T_{\rm max}$ at all. This is because the RG 
trajectories for $\lambda(\mu)$ run almost horizontally at high energies, and  
 touch the critical line of stability $\lambda_c(T)$ at the bend at 
$T=T_R$ (see Figure \ref{fig:criticalT}). 
Vacuum decay becomes most effective at the epoch of reheating in this case, and it does not matter 
how high $T_{\rm max}$ is. The fact that the RG trajectories for $\lambda$ run 
almost horizontally at high energies also explains 
why the prediction for $m_H$ levels off and depends
only weakly even on $T_R$ 
at high reheating temperatures. The Higgs quartic coupling does not run much at 
 high energies because of the smaller top Yukawa coupling.

Figure \ref{fig:criticalT}a shows that the RG trajectory corresponding 
to the prediction of section 2 (the lowest trajectory) crosses $\lambda_c(T)$ at around 
$\mu = 10^8$ GeV. Thus, the prediction of section 2
is unaffected if $T_R = T_{\rm max} \simlt 10^8$ GeV in the instantaneous 
reheating scenario, as shown by point $P_2$ in Figure
\ref{fig:mhprediction}. 
The highest reheating temperature for which 
the prediction of section 2 persists actually depends on the relation between
$T_R$ and $T_{\rm max}$. The higher $T_{\rm max}$ is, the lower the maximum value of $T_R$ for which
the vacuum prediction applies.  In the limit  $T_{\rm max} \gg T_R$,
this value of $T_R$ drops to around $10^5$ GeV,  as shown by point $P_1$ in Figure
\ref{fig:mhprediction}.

Finally, we have a few remarks about the uncertainties associated with 
the calculations of this section. The landscape prediction (\ref{eq:pred-val}) in section 2, 
was derived from a value
of the 
Higgs quartic coupling at a very high energy scale, $M_{dom}$. 
The three sources of uncertainties quoted in (\ref{eq:pred-val}), namely, 
the measured top quark mass, QCD coupling  and calculation of 
 electroweak threshold corrections, potentially changed the relation between 
$\lambda(M_{dom})$ and $m_H$. 
The prediction in Figure \ref{fig:mhprediction} from the thermal scenario came from couplings 
 renormalized at a lower energy scale $\sim T \ll M_{dom}$.
However, the running of  $\lambda(\mu)$ takes place mainly 
at low energy, where the top Yukawa coupling is large.  
Thus, the prediction for the Higgs mass at high reheating temperature  is also subject to the same
three uncertainties quoted in (\ref{eq:pred-val}) with almost equal sizes.
The influence of the details of the 1-loop functional determinant was 
at most 1 GeV \cite{Strumia} for the prediction in (\ref{eq:pred-val}), 
and it will remain at that order of magnitude in Figure \ref{fig:mhprediction}
as well, because the estimates of the 1-loop functional determinant are 
much the same; see footnote \ref{fn:det}.

It should be noted that the prediction in the case of high reheating temperature is associated 
with an additional uncertainty: the precision of the calculation 
of the thermal effective potential. This issue is briefly discussed 
in the appendix.

\subsection{The Width of the High Temperature Prediction}

The width of the Higgs mass prediction for this section may be obtained similarly to 
that of section 2.  We consider the total probability ${\cal P} = P f$, and again we may
define $\lambda_{-}$ and $\lambda_{+}$ to be the values of the quartic
coupling at $\Lambda_{SM}$ for which ${\cal P}$ has fallen from its peak
value by a factor of 1/e. Since the cut-off from $f$ is so sharp, 
$\lambda_{-}$ is again a negligible distance from the peak.  
As in section 2, $\lambda_{+}$ is shifted above the critical value
$\lambda_c$ by an amount $|\lambda_c|/p$, where here 
we are using the value of the critical quartic coupling run up to the
scale $\Lambda_{SM}$.  $p$ is defined as in section 2 by
 $ p = \left. \frac{\partial \ln P(\lambda_0)}{ \partial \ln \lambda_0} 
\right|_{\lambda_0 = \lambda_{c}}$.
This yields a possible upwards shift in the Higgs mass of 
\begin{equation}
\Delta m_H \approx \frac{550|\lambda_c|\GEV}{p}.
\end{equation}
Here, $|\lambda_c|$ ranges from about .02 to .05 depending on the
particular thermal history after inflation. 
This gives an upwards shift between $(\sim 10 \, \GEV)/p$ and  $(\sim 25
\, \GEV)/p$.

\subsection{Discussion}

In the case that SM patches of the universe were reheated to very high 
temperatures after inflation, our prediction for the central value of
the Higgs boson mass is shown in Figure \ref{fig:mhprediction}. 
Despite a very wide range in the cosmologies considered, the central
value remains in the region (106--118) GeV. Furthermore, for $T_R < 10^5$
GeV thermal fluctuations are irrelevant and the 106 GeV result from
quantum tunneling applies, while for $T_R > 10^{12}$ GeV, the
prediction is in the narrow range $(117 \pm 1)$ GeV.
The uncertainties from the
experimental values of $m_t$ and $\alpha_s$ and from higher loop
orders are closely similar to those in the non-thermal case. The peaking
of the a priori distribution for $\lambda_0$ yields an upper width on the prediction
that is higher than the central value by an amount between $\sim 10 \, \GEV /p$ and 
$\sim25 \, \GEV /p$, depending on the reheat temperature.

The top mass prediction of section 3 is very significant. If the top
Yukawa coupling $h_0$ is allowed to scan, does this
prediction survive thermal fluctuations at high temperatures? 
If $T_R > 10^8$ GeV, then a large $p$ will causes $h_0$ to
rise significantly above the observed value, since a larger $h_0$
gives both a steeper quartic trajectory and a lower $\lambda_c(T)$
curve.  
Thus, to maintain the highly significant $m_t$ prediction from section
3, we are led to a preference for low $T_R$.

\section{The Scanning Standard Model}

Consider the SM, minimally augmented with right-handed
neutrinos to allow for both neutrino masses and leptogenesis, with all
parameters scanning---the scanning SM.   
It is possible that hospitable parts of
the landscape exist where parameters take on values that
are very different from the ones we observe.  This is well-illustrated
by the weakless universe \cite{Harnik:2006vj}, where $\vev{H}/M_{Pl} \approx 1$ while the
Yukawa couplings for the light quarks and the electron, $h_{u,d,s,e}$,
are all less than $10^{-20}$.  We assume that, if such regions exist,
they are less probable than our own universe.  For example, the
weakless universe may be disfavored because of a very small
probability to have four Yukawa couplings that are so small.  In the
landscape, the origin for $v \ll M_{Pl}$ may be that a small value
for $v$ is the most cost effective way of keeping $u,d,s$ and $e$ light.
Hence the question becomes: in our neighborhood of the landscape, how
much of the parameter space of the scanning SM can be understood from
environmental selection?  We have argued that the entire Higgs
potential is strongly selected, and now we turn to the Yukawa coupling
matrices, where most of the parameters lie.
In sections 2, 3, and 4 we have been exclusively focused on precise
predictions for the Higgs and top masses. In this section we explore
possible consequences of embedding our previous results into the
scanning SM.  Further assumptions are found to lead to intriguing
order of magnitude estimates for a variety of quantities.

\begin{figure}
\begin{center}
\includegraphics[width=1.0\linewidth]{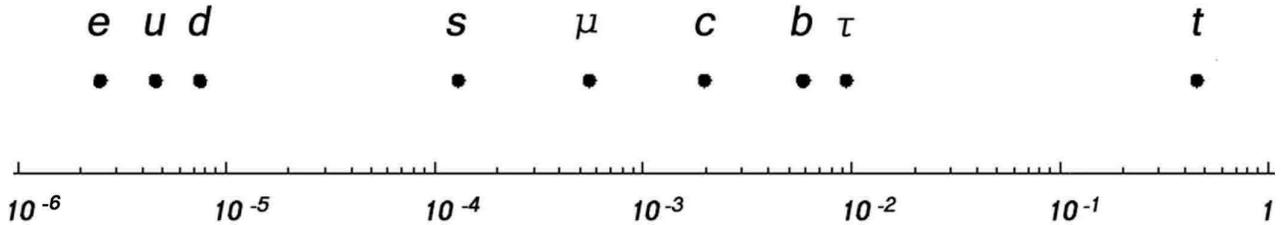}
\caption{\label{fig:yukawas} The Yukawa couplings for the quarks and charged 
leptons scaled up to $10^{18} \, \GEV$, and plotted logarithmically.  
The environmentally important $e, u, d$ and $t$ couplings seem to be outliers, 
while the remaining five couplings are all clustered in a two order of magnitude 
range.}
\end{center}
\end{figure}

\subsection{Quark and Charged Lepton Masses}

Let us first ignore flavor violation and
consider scanning of the nine charged fermion eigenvalues.
In Figure \ref{fig:yukawas} we plot the nine Yukawa couplings of the
SM at the renormalization scale $\Lambda_{SM} = 10^{18}$ GeV.  
They fall into three well separated groups:  the $e,u$ and $d$ Yukawas,
selected to be small by the requirement that atoms exist \cite{barrowetal},
the $t$ coupling, selected to be large to avoid vacuum
instability, and the remaining five Yukawa couplings that are
environmentally irrelevant, and therefore reflect the underlying probability
distribution on the landscape. It is striking that while the nine
couplings span 5 orders of magnitude, the five irrelevant couplings
span only 2 orders of magnitude.  It therefore seems possible to make
progress on the fermion mass hierarchy problem by assuming a universal
fermion Yukawa eigenvalue probability distribution that is peaked
around $10^{-3}$, with a width of one to two orders of magnitude. Possible
simple analytic forms for the distribution include 
\begin{equation}
P(\tilde{h}) = A \, e^{- \left( \frac{\log_{10} \tilde{h}}{\sqrt{2}\sigma} \right)^2}
\label{eq:Pyuks1}
\end{equation}
with $\tilde{h} = h_0/10^{-3}$ and $\sigma \simeq 0.75$, and  
\begin{equation}
P(\tilde{h}) = A \, \tilde{h} \, e^{- \tilde{h}}.
\label{eq:Pyuks2}
\end{equation}
Thus the five irrelevant Yukawa couplings
are scattered logarithmically about $10^{-3}$, while the four outliers are
environmentally selected to be far from the peak of the
distribution.\footnote{With these distributions, the weakless universe is
indeed much less probable than the observed SM.} This interpretation
of the data of Figure \ref{fig:yukawas} seems so plausible, that we
investigate further the role of electroweak phase stability
in forcing the top coupling to large values.

From Figure \ref{fig:running4} it is clear that if a strongly peaked distribution
$P(\lambda_0)$ pushes $\lambda_0$ towards its critical value
$\lambda_{0,c}$, then vacuum instability is only avoided if there is a
Yukawa coupling that is sufficiently large to RG scale $\lambda$ to
positive values by the weak scale.  This can be accomplished by a
large top coupling via the term  $-6 h_t^4$ of (\ref{eq:lambdarge}).
However, with all Yukawa couplings scanning, the $-6 h_t^4$ term could
be  replaced by other large Yukawa contributions. One easily sees
that it is more probable to have a single large Yukawa rather than two
or more large Yukawas making significant contributions to the quartic
RG equations. Furthermore, it is not sufficient for the single large Yukawa to
be that of a lepton: it is only for a quark that QCD increases the
Yukawa in the infrared, thus magnifying the effect on the scalar
quartic coupling.  The critical value for a lepton Yukawa coupling is slightly 
larger than for a quark, but this is insufficient to allow
a critical quartic coupling at $\Lambda_{SM}$ to become positive by
the weak scale.  Hence, our simple interpretation of
Figure \ref{fig:yukawas}, in terms of a universal $P(h_0)$ and
environmental selection of outliers, predicts not only that there will
be a single heavy fermion, but that it must be a quark. However, the choice of
up-like versus down-like is random.  Crucially, as we have shown in
section 3, a precise and successful prediction for the top quark 
results.\footnote{Note that the distribution (\ref{eq:Pyuks1}) gives a
  peaking parameter $q \simeq 2$.  While this has the right sign to
  limit $\Delta m_{t+}$, the size of $\Delta m_{t+}$ from (\ref{eq:mt+}) seems too
  large. Hence, either $P(h)$ drops more rapidly at large $h$ compared
to  (\ref{eq:Pyuks1}), or  $\Delta m_{t+}$ is made small by a large
tunneling rate in the full theory above $\Lambda_{SM}$. 
}

If this picture of fermion masses is correct, then there is a further 
consequence for the $u,d$ and $e$ masses.  Since they are selected to be
quite far from their most probable values in the landscape, our patch of 
the universe is quite rare. This means that they are expected to be quite 
close to the maximum values consistent with environmental selection for 
the existence of complex atoms.  
Of course, this is sensitive to the actual probability distribution
for the Yukawa couplings.
For example, using the distribution 
(\ref{eq:Pyuks1}) patches in the multiverse with the electron Yukawa double 
its observed value are more than an order of magnitude more probable 
than in our patch.  On the other hand, for the distribution 
(\ref{eq:Pyuks2}) they are only twice as probable.
 
It is not clear how the small CKM mixing angles could arise from
environmental selection.  Vacuum stability has selected an up-type
quark to be heavy, and nuclear physics has selected a down-type quark
to be light. In nature these two quarks are almost precisely in
different weak doublets, so that $V_{td} \simeq 10^{-3}$, but there does not
appear to be a selection effect to explain this.  This is similar, 
but numerically more severe, to the puzzle of why environmental 
selection would make $V_{ud}$ close to unity. Quark flavor violation 
appears to be governed, at least to some degree, by symmetries.  
Numerical simulations of Yukawa matrices have been made for very flat
landscape probability distributions  \cite{Donoghue:2005cf}.
It would be interesting to perform numerical simulations using more
peaked distributions, such as (\ref{eq:Pyuks1}) or (\ref{eq:Pyuks2}),
and incorporating elements of approximate flavor symmetries. 

\subsection{Neutrino Masses, Leptogenesis and Inflation}
Since no large neutrino mass ratios or very small neutrino mixing
angles have been measured, it is very plausible that the entries of the neutrino mass matrices,
both Dirac and Majorana, are distributed according to some universal
probability distribution of the landscape. Indeed, Neutrino Anarchy demonstrated in
some detail how randomly generated neutrino mass matrices could
account well for the observed masses and mixings \cite{Hall:1999sn}.
The view of the charged fermion masses presented above has an important 
consequence for neutrino physics. We assume that 
the Dirac neutrino Yukawa couplings are governed by the same landscape 
probability distribution as the quarks and charged leptons. 
If there is little environmental selection acting on 
the neutrinos, then the typical Dirac Yukawa eigenvalue $\overline{h}_\nu$ is expected to be 
of order $10^{-3}$ in magnitude.\footnote{Sufficiently small that these couplings do not upset the 
prediction for the Higgs boson and top quark masses.} Since no large hierarchy of 
neutrino masses is observed, it is reasonable to also assume that the 
three Majorana masses of the right-handed neutrinos are governed by a 
universal probability distribution, so that they are all expected 
to have a common order of magnitude,  $\overline{M}_R$. With 
$\overline{h}_\nu \approx 10^{-3}$, there is a relation between 
$\overline{M}_R$ and the order of magnitude of the light neutrinos, 
$\overline{m}_\nu$:
\begin{equation}
\frac{\overline{m}_\nu}{\mbox{eV}} \approx
\frac{ 3 \times 10^7 \, \GEV}{\overline{M}_R}
  \left( \frac{\overline{h}_\nu}{10^{-3}} \right)^2
\label{eq:MRmnu}
\end{equation}
In our patch of the multiverse, $\overline{m}_\nu$ has been observed to be of order 
$10^{-2}$ eV, so that we predict 
\begin{equation}
 \overline{M}_R \approx 3 \times 10^9 \, \GEV
\label{eq:MR}
\end{equation}
in our patch.

Leptogenesis is the only possible source of a baryon asymmetry in the scanning 
SM. Is it consistent with our framework for fermion masses? 
The lepton asymmetry is given by \cite{Buch}
\begin{equation}
Y_L \approx \kappa \frac{\epsilon_1}{g_*},
\end{equation}
where $g_* \simeq 100$ counts the number of states in the thermal bath.
The CP asymmetry $\epsilon_1$ and the washout factor $\kappa$ are roughly 
of order 
\begin{equation}
  \epsilon_1 \sim \frac{3}{16 \pi}\bar{h}^2_\nu \sim 10^{-1} \bar{h}^2_\nu,
  \qquad
  \kappa \sim 10 \frac{1}{\bar{h}^2_\nu}\frac{\overline{M}_R}{M_{Pl}},
\end{equation} 
respectively, where we have assumed comparable masses for the right-handed 
neutrinos.
We are now assuming that all relevant entries of the neutrino Yukawa 
coupling matrix are of order $\overline{h}_\nu$. 
We see that the observed baryon asymmetry is roughly reproduced after
using (\ref{eq:MR}):
\begin{equation}
  Y_L \approx 10^{-11} \left( \frac{\overline{M}_R}{3 \times 10^9 \, \GEV}
     \right) \approx 10^{-11} \left(\frac{\overline{h}_\nu}{10^{-3}}\right)^2
    \left(\frac{10^{-2} \, \EV}{m_\nu} \right).
\label{eq:YL}
\end{equation}
This can be regarded as an indication that the Dirac Yukawa couplings of 
neutrinos are also subject to the universal distribution $P(h_0)$ 
of Yukawa couplings.

The baryon asymmetry is roughly derived 
from the observed low-energy neutrino masses, or vice versa; 
either one of them can be used as an input so that the other is determined.
Now the next question is whether we can understand them all together 
environmentally. Suppose that the Majorana masses of the right-handed 
neutrinos are scanned. The distribution of $M_R$ may be a 
featureless flat distribution, like that of the cosmological constant; 
or, the baryon asymmetry, being proportional to $M_R$ (as in (\ref{eq:YL})), may put more 
weight on larger values of $M_R$. We assume that the distribution of $M_R$ 
is mildly peaked toward larger values. If the reheating temperature $T_R$ 
is also scanned mildly, but with an a priori distribution peaking weaker 
than that of $M_R$, then both $M_R$ and $T_R$ would go hand in hand, 
pushed upward, because the baryon asymmetry is at least power-suppressed 
when $T_R$ is much lower than $M_R$.
Both $M_R$ and $T_R$ go up until $T_R$ and $T_{\rm max}$ are so high 
that the Higgs vacuum becomes unstable. If $T_{\rm max}$ is also scanned, 
with an even milder a priori distribution, $T_R$ will be pushed upward 
along the line from $P_1$ to $P_2$ in Figure \ref{fig:mhprediction} 
until it reaches the special point $P_2$. For a large $p$, on the 
other hand, a reheating temperature higher than that is less probable, 
because a large $p$ strongly favors a smaller $\lambda_0$, which 
is environmentally allowed only in patches with lower $T_R$.
Thus environmental selection for the desired electroweak vacuum, 
on the combined a priori probability distribution, leads to the special point 
$P_2$ as the most likely one to be observed, and hence to the predictions 
\begin{equation}
T_R \approx T_{\rm max} \approx 10^8 \, \GEV. 
\label{eq:TR}
\end{equation}
The reheat temperature is then low enough that the predictions for the Higgs and 
Top masses given in sections 3 and 4 remain valid.

Since $T_{\rm max}^4 \approx \sqrt{T_R^4 V_I}$, we predict the vacuum
energy during inflation 
\begin{equation}
V_I \approx (10^8 \, \GEV)^4 
\label{eq:rhoI}
\end{equation}
corresponding to a Hubble parameter during inflation of order
$10^{-3}$ GeV.  These speculations will clearly be disproved if a tensor
perturbation is seen in the cosmic microwave radiation.  From the
reheat temperature we predict the decay rate of the inflaton
\begin{equation}
\Gamma_I \approx 10^{-3} \, \GEV. 
\label{eq:gammaI}
\end{equation}
Now that the rough scale of $M_R$ is obtained without an observational 
input, low-energy neutrino masses of order $10^{-2} \, \EV$ and 
the baryon asymmetry are also purely predictions of 
environmental selection.

Since in thermal leptogenesis the lightest right-handed neutrino mass can 
be slightly higher than the reheating temperature $T_R$, 
there is no conflict between $T_R$ in (\ref{eq:TR}) and $\overline{M}_R$
in (\ref{eq:MR}).   Also, the neutrino Yukawa couplings contributing 
to the CP asymmetry $\epsilon_1$ may fluctuate bigger, and those appearing 
in the denominator of the washout factor $\kappa$ smaller, making the 
entire baryon asymmetry bigger. Thus an apparent discrepancy of one order of 
magnitude is nothing to worry about. Although the reheat temperature (\ref{eq:TR}) 
does not satisfy the phenomenological limit $2 \times 10^9 \, \GEV < T_R$
derived from the observed baryon asymmetry \cite{Buch}, our estimates are only order of 
magnitude ones.  Furthermore, this limit does not apply 
when the right-handed neutrinos have highly degenerate Majorana masses.
An environmental factor $\propto Y_B$ may motivate 
such highly degenerate right-handed neutrinos.

\subsection{Gauge Couplings}

If all three SM gauge couplings scan, then $\alpha$ and $\alpha_s$ are
selected, while the weak mixing angle $\sin^2 \theta$ is
apparently environmentally irrelevant. However, the condition for the critical value
for the top quark coupling, equation (\ref{eq:hc1}), depends on the weak mixing
angle. Hence it is possible that avoiding vacuum instability could
select for the weak mixing angle in addition to the top quark and
Higgs masses!  If the a priori probability distribution has a
stronger dependence on $h_t$ than $\sin^2 \theta$ then we find the incorrect
prediction of $\sin^2 \theta \simeq 0.38$ at the weak scale.  Hence
it must be that the weak mixing angle is essentially determined by
the landscape distribution.

Gauge coupling unification is possible in the scanning SM, with 
$\Lambda_{SM}$ near $M_{Pl}$, if threshold corrections are of 
order 10--20\%. Since the threshold corrections to the gauge coupling 
constants arise from dimension-5 operators, while the leading order
correction to the Higgs potential (\ref{eq:quartic-pot}) is a dimension-6 operator, 
corrections to gauge coupling unification can be as large as 10--20\% 
without losing the 1\%-level precision of the predictions in earlier sections.

\subsection{Dark Matter}

The final parameter of the SM that could scan is the strong CP
parameter of QCD.  It is attractive to promote
this to the axion field, with an associated symmetry breaking scale above
$\Lambda_{SM}$. It has been argued that scanning the primordial value
of this axion field not only allows axionic dark matter, but could
explain the approximate equality of dark matter and baryon energy
densities \cite{Wil}. 

\vspace{0.2in}

The scanning SM is frustrating in the sense that there appears little
that is amenable to experimentation beyond the Higgs mass prediction.
However, the question of flavor in both quark and neutrino sectors is left
unresolved, and in the neutrino sector more thought needs to be given
to the environmental selection effects arising from leptogenesis.
 
We caution the reader that there are potential difficulties with the
scanning SM that we do not address in this paper.  For example, particle
physics of the SM is completely unchanged if $m^2$ and the QCD scale
$\Lambda_{QCD}^2$ are increased by the same factor, and corresponding
changes are made to the dimensionless couplings so that their values 
at the scale $m^2$ are kept fixed.  The environmental selection
of the weak scale to be as small as 100 GeV would then be undermined, 
unless some probability distribution can be found to counteract 
the one favoring large $m^2$, or some other environmental selection 
involving the Planck scale, such as big bang nucleosynethesis, 
limits such a scanning direction.
Similar questions arise for the environmental selection of the
cosmological constant when there is simultaneous scanning 
of other cosmological quantities, such as the density perturbations.  
This has been addressed by a consideration  of other selection effects \cite{TR}.

\section{Conclusions and Discussion}
 
There are many mass scales that play an important role in physics, 
for example the electron and proton masses, but from a fundamental
viewpoint two are key:  the scale of electroweak symmetry breaking,
$\vev{H} = v$, that sets the mass scale of the quarks and leptons as well as the
$W$ and $Z$ bosons, and the Planck mass, $M_{Pl}$, that describes the
gravitational coupling and is the largest scale that enters
physics. We take the viewpoint that the Planck scale is fundamental
and the weak scale is somehow to be derived, although the alternative
is also possible, so that the key question becomes the origin of the
small dimensionless parameter $v/M_{Pl}$. 

Over the last 30 years, many frameworks have been developed that can
explain this small ratio by introducing various symmetries.
The LHC is expected to play a crucial role in distinguishing  between
these frameworks, for example between supersymmetry and new strong forces 
at the TeV scale.  The alternative viewpoint, that $v/M_{Pl}$ is
environmentally selected \cite{Agrawal:1997gf} and has nothing to do with extended
symmetries, has received relatively little attention, perhaps because
it leads to no new physics, beyond a light Higgs boson, to be
discovered at LHC.  Of course it is possible that in this case the LHC
might discover physics unrelated to electroweak symmetry breaking,
perhaps associated with either dark matter or coupling constant
unification, but it is hardly to be expected.  
Dark matter is not necessarily related to the 
electroweak scale; axion dark matter with an environmentally 
chosen relic density is an example. Gauge coupling unification 
requires new particles other than a Higgs doublet, but those 
particles may be around the Planck scale, for instance. If the 
origin of $v/M_{Pl}$ is environmental, then arguments for physics
discoveries beyond the SM at LHC become tenuous.

Another fundamental mass scale of nature is the cosmological constant,
$\Lambda^4$. To date, theories beyond the SM have failed to provide 
a symmetry understanding for why $\Lambda/M_{Pl}$ is so small: 
why it is so close to zero, and what determines the order of 
magnitude of the deviation from zero. A symmetry might assign 
a special meaning to a vanishing cosmological constant, and it 
might be possible to understand the observed size of the dark 
energy by constructing theories that implement a seesaw
relation $\Lambda \approx v^2 /M_{Pl}$. 
On the other hand, environmental selection explains  
{\em both} the extreme smallness of $\Lambda$ {\em and} its order of
magnitude, as measured by the dark energy density.
The dark energy 
and the ratio $v/M_{Pl}$ are environmentally selected if two 
parameters of the renormalizable Higgs quartic potential are
scanned. These successes motivate us to think of scanning 
the entire Higgs potential. In this case, the Higgs quartic 
coupling, $\lambda$, and therefore the physical Higgs mass,
scans. By deriving a prediction for the Higgs boson mass, 
the idea of the scanning Higgs potential can be tested.

Environmental selection requires an ``edge'' \cite{ADK,Aguirre}--- 
a surface in the space of
scanning parameters, such that on one side of the surface the
formation of desired complex structures is greatly suppressed.  We do
not know enough about conceivable life forms to derive the precise
types of complexity to be selected.  However, the more general the
requirement, the more plausible selection becomes. The selection of
small values for $\Lambda / M_{Pl}$ results from an edge that
corresponds to formation of non-linear structures late in the universe.
As $\Lambda/M_{Pl}$ is increased beyond this edge, the probability of
such structures forming is suppressed because they are subjected to
inflation before they go non-linear. On the other hand, 
a small value for $v/ M_{Pl}$ is selected by the requirement that
atoms exist. The requirement that atoms exist is much more specific than the
requirement that large non-linear structures form, but it is not
unreasonable.  For our prediction of the Higgs boson mass, the
relevant edge is the phase boundary that separates a metastable phase
with small $\vev{H}/M_{Pl} = v/M_{Pl}$ from a phase with 
$\vev{H}/M_{Pl} \approx 1$. This is less
specific than requiring the existence of atoms: life probably requires
objects that contain a large amount of information, and this may
be accomplished best in a phase with a large $M_{Pl}/v$ ratio. 

An edge that allows environmental selection is not sufficient to make
predictions; some knowledge of the a priori probability distribution
for the scanning parameters is needed.  In the absence of a
calculation from a fundamental theory of the landscape, this requires
an assumption.  For the cosmological constant and weak scales, this
assumption is extremely mild; it is sufficient to assume that the
distributions are roughly flat and featureless, so that small values of these
scales are exceedingly rare. For our Higgs mass prediction a
non-trivial assumption is necessary: the probability distribution
$P(\lambda)$ must be sufficiently peaked at low coupling
that $\lambda$ is expected to be near the metastability boundary.

Since our edge is a phase boundary, the position of the
boundary, and therefore the Higgs mass prediction, depends on the
thermal history of the universe, in particular on the reheat and
maximum temperatures
after inflation, $T_R$ and $T_{\rm max}$,  two of the most important
parameters of post inflationary cosmology that are still largely 
unconstrained, and consequently one might expect that little 
can be said about the Higgs mass.  
In fact, no matter what the values of these two temperatures the most
probable value of the Higgs mass is raised from 106 GeV to at most 118
GeV. 
Furthermore, if $T_{\rm max} = T_R \simlt 10^8$ GeV
the prediction is very insensitive
to cosmology, as the danger of a phase
transition is highest today and was negligible in the early
universe, giving a central value
\begin{equation}
\overline{m}_H =  106 \, \GEV  + 6 \, \GEV  \left( \frac{m_t - 171 \, \GEV}{2 \, \GEV}
\right) - 2.6 \, \GEV \left(\frac{\alpha_s - 0.1176}{0.002}\right)
\pm 6 \, \GEV . 
\label{eq:mH1}
\end{equation}
%
%
%
with an upward statistical fluctuation of (25/p) Gev.  For $T_R > 10^{12} \, \GEV$, nucleation at high temperature
dominates, giving $m_H = 117 \pm 1 \, \GEV$ for any value of
$T_{\rm max}$, with an upper fluctuation between $\sim (10/p)$ GeV  and $\sim (25/p)$ GeV from
the a priori distribution, depending on the reheat temperature.
The full dependence of the prediction on $T_R$ and $T_{\rm max}$ is shown in Figure
\ref{fig:mhprediction}.  The parameter $p$ is the logarithmic derivative of $P(\lambda)$
evaluated at the phase boundary.  Further uncertainties, displayed
in equations (\ref{eq:pred-val}) and (\ref{eq:DeltamH}), arise from the
experimental uncertainty in the QCD coupling and the top
quark mass, and also from use of perturbation theory in RG scaling
of couplings and in relating couplings to physical masses.
Ultimately, these uncertainties can be reduced; but
the width of the Higgs boson mass distribution from the
multiverse, parameterized by $p$, cannot be reduced.
This width is comparable to the present uncertainties if $p \approx 3$, 
while if $p > 10$ it is practically negligible compared
with other uncertainties.

The top quark plays a crucial role in electroweak symmetry breaking
via the RG evolution of the quartic coupling $\lambda$.  If the top
Yukawa coupling also scans, a further mild assumption also allows a
prediction for the top quark mass. For $T_R < 10^8$ GeV, 
the most probable top quark mass is
\begin{equation}
m_t = \left[176.2 \pm 3 + 2.2 \log_{10}(\Lambda_{SM}/ 10^{18} \, \GEV) \right] \, \GEV.
\label{eq:mtfinal}
\end{equation}
To limit fluctuations above this requires a further assumption, while
downward fluctuations are $35 \, \GEV / \sqrt{p}$.
The striking success of this result
suggests that environmental selection may be at work in the Higgs
potential, and that we may be lucky enough for $p$ to be sufficiently large
to give a precise Higgs mass prediction.
Maintaining this prediction in the presence of thermal fluctuations also
implies that $T_R \simlt 10^8$ GeV.

We have argued that these Higgs and top mass predictions could occur 
in the scanning SM, where all SM parameters scan.  Indeed a
zeroth-order understanding of the charged fermion masses follows if
all Yukawa couplings have a universal probability distribution
centered on about $10^{-3}$.  An interesting extension to neutrino
masses and leptogenesis follows if right-handed neutrino masses scan, with a 
preference for larger values, and if $T_R$ and $T_{\rm max}$ scan with mild
distributions.  The broad order of magnitude of the light neutrino
masses and the baryon asymmetry are correctly predicted, 
while the right-handed neutrino masses, the reheat temperature and 
the maximum temperature are all predicted to be of order $10^8\mbox{--}10^9$ GeV.

The possibility of precise predictions for the Higgs and top masses by
environmental selection from a landscape illustrates that it
may be possible to do physics without symmetries. 
Theories based on symmetries yield predictions because the symmetries
limit the number of free parameters. One chooses a highly symmetric
model and studies the resulting predictions.  
On the landscape two ingredients are needed for a
``bottom-up'' prediction:  an edge that allows environmental
selection,  and an assumption that the a priori probability
distribution is pushing parameters towards this edge.  
Predictions follow, quite literally, from living on the edge.  Thus inventing
models is replaced by finding edges and inventing probability
distributions.  There are two difficulties with this approach:
relevant edges are hard to come by, and even if environmental
selection is occurring at an edge, we can only discover it if the
probability distribution happens to be favorable.

If the LHC discovers a light Higgs boson in our predicted range, and
no sign of any physics beyond the SM, then a precise numerical test of
environmental selection could follow from further developments in both
theory and experiment. This will require a reduction in the experimental error bars of
the QCD coupling and the top quark mass, and improved accuracy of
calculations of both RG scaling and extracting pole masses from
running couplings.  In the case that the phase boundary is determined
by thermal fluctuations at high temperatures, a refined calculation of the
temperature dependent effective potential will be needed, in
particular including the effects of Higgs quanta in the thermal bath.

\vspace{5mm}

We conclude by addressing those who are sceptical of physics arguments 
from environmental selection.  We have shown how precise predictions 
for the Higgs and top masses arise from assumptions about the landscape. 
If these precise predictions are found to be false, then our theoretical 
construct will be experimentally disproved. If they are found to be
correct, then the SM is indeed on the edge of electroweak vacuum instability. 
It would then be up to you to find an alternative, more convincing
theoretical framework to explain this fact.   
In this situation, we suspect that effort will be focussed on extracting 
further predictions from the landscape,
and on understanding how the landscape might arise from string theory.

\section*{Acknowledgments}  

We thank A. Strumia for communications.
This work was supported in part by the Director, Office of Science, 
Office of High Energy and Nuclear Physics, of the US Department of 
Energy under Contract DE-AC03-76SF00098, in part by the National
Science Foundation under grant PHY-04-57315 and in part by 
Miller Institute for Basic Research in Science. 
LH and TW thank the Aspen Center for Physics where this work was completed.

\appendix 

\section{A Little More about The Thermal Effective Potential}

At the 1-loop level, the effective potential in a thermal bath $V_{\rm tot}(H)$
can be split into a sum of the effective potential at $T=0$ and a
thermal 1-loop contribution, $V_{TH}(H;T)$. The thermal contribution
is given by \cite{Anderson, AV, Sher, EQ}
\begin{equation}
V_{TH}(H; T) =\frac{T^4}{2\pi^2} \left( 6 J_B \left( \frac{g_L}{2} \frac{H}{T} \right)
   + 3 J_B \left(\frac{\sqrt{g_L^2 + g_Y^2 }}{2} \frac{H}{T} \right)
   - 12 J_F \left( \frac{h}{\sqrt{2}} \frac{H}{T} \right) \right),
\label{eq:WZT}
\end{equation}
where
\begin{eqnarray}
 J_B(y) & = & \int_0^\infty dx x^2 \ln \left( 1 - e^{-\sqrt{x^2+y^2}} \right), \\
 J_F(y) & = & \int_0^\infty dx x^2 \ln \left( 1 + e^{-\sqrt{x^2+y^2}} \right).
\end{eqnarray}
Here, only contributions of loops of W-bosons, Z-bosons and top quarks are 
taken into account, and not of the Higgs field itself. 
For Higgs field values much less than the temperature, this 
thermal contribution to the effective potential is
approximated by a free energy piece plus quadratic term.
On the other hand, for $T \ll H$, $V_{TH}(H;T)$ in (\ref{eq:WZT}) 
is exponentially small.

\begin{figure}
\begin{center}
\begin{tabular}{ccc}
\includegraphics[width=.3\linewidth]{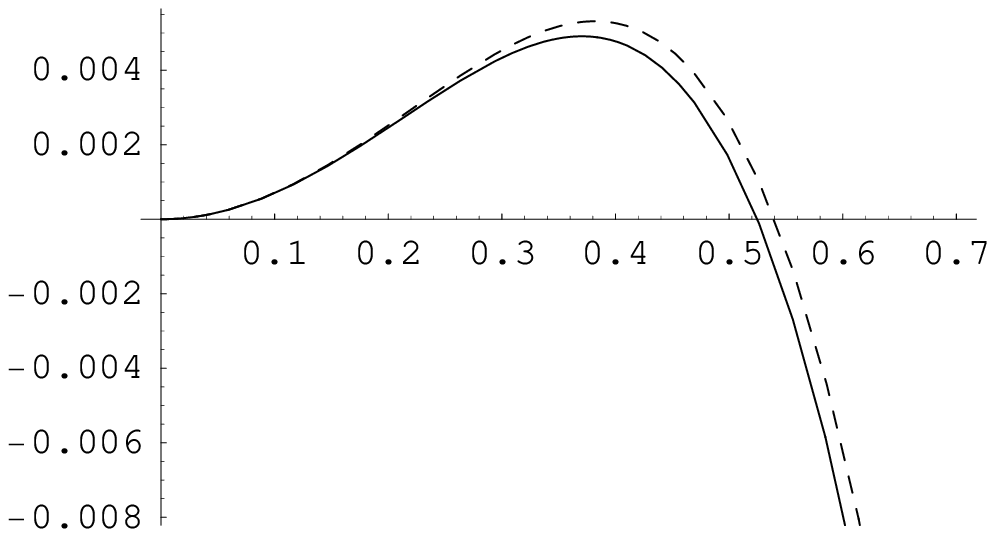} &
\includegraphics[width=.3\linewidth]{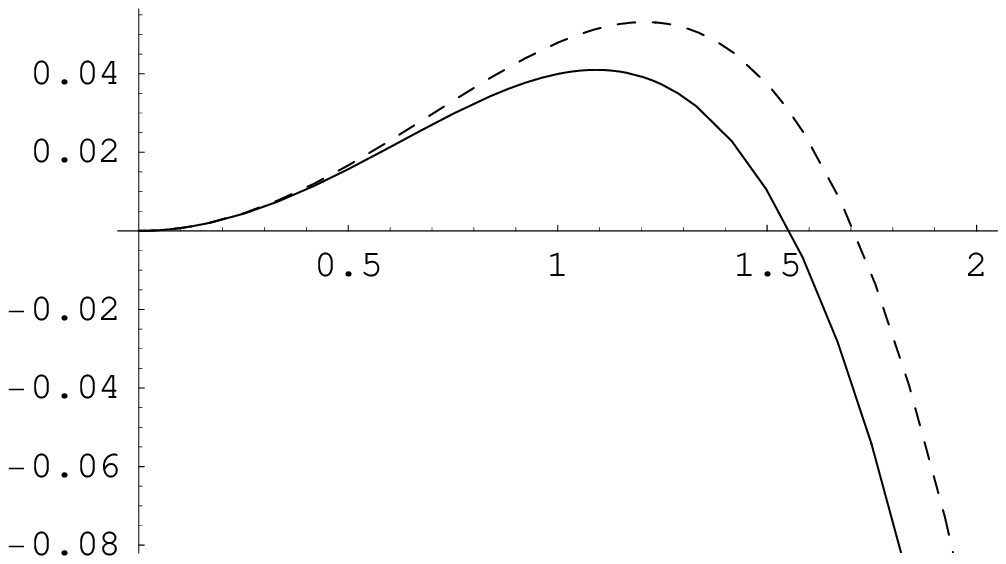} &
\includegraphics[width=.3\linewidth]{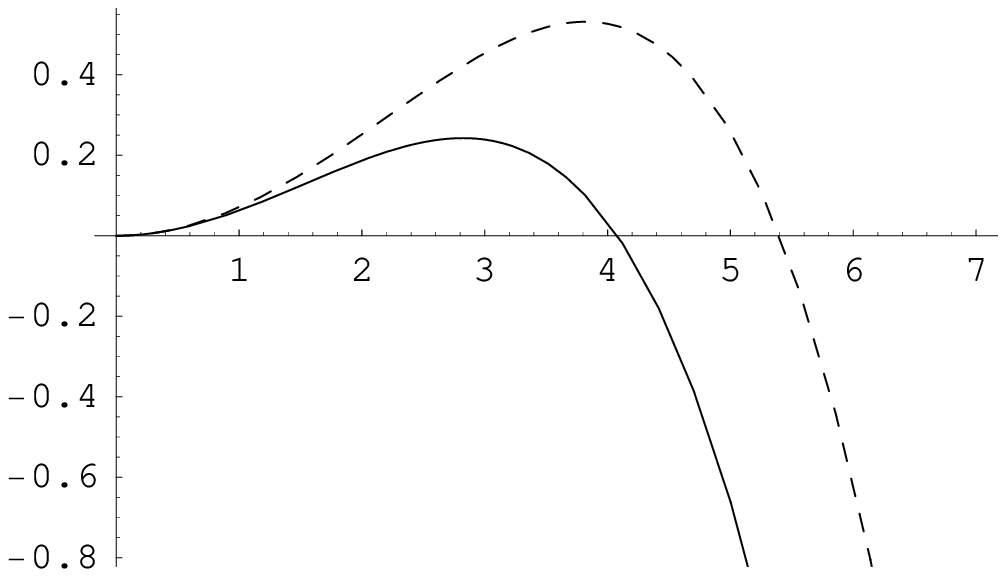}
\end{tabular}
\begin{picture}(0,0)(450,50)
%
%
\Text(165,95)[b]{$V_{\rm tot} / T^4$}
\Text(280,65)[l]{$\frac{H}{T}$}
\end{picture}
\caption{\label{fig:thrmpot} Effective potential in the thermal bath
 $V_{\rm tot}(H)$
for different values of $\lambda$, (from left) $-1$, $-0.1$ and $-0.01$.
The potential with (\ref{eq:WZT}) is drawn by solid lines, and with its approximation
(\ref{eq:thmass}) by dashed lines. Using $g_L = 0.570$, $g_Y = 0.402$ and
$h = 0.547$, we find $g_{\rm eff} \simeq 0.382$ and $g_{\rm eff}^4 \sim
 0.02$ for $T = 10^{10}$ GeV and $m_H = 120$ GeV.}
\end{center}
\end{figure}

For a negative Higgs quartic potential with large $|\lambda(T)|$,
the potential barrier for the vacuum transition is at a small field value
of $H$. Hence, the overall shape of the potential is almost properly obtained
even if the thermal contribution is approximated by the quadratic term
(see Figure \ref{fig:thrmpot}a). On the other hand, for a small $|\lambda(T)|$,
the potential barrier is outside the range where the high-temperature expansion
is valid, and $V_{\rm tot}$ calculated using the quadratic approximation for the
thermal contribution (\ref{eq:thmass}) is quite different from the
right shape of the potential (see Figure \ref{fig:thrmpot}c).
The quadratic approximation $V_{TH}^{(2)}(H ; T)$ in (\ref{eq:thmass}) continues to grow
for large $H$, but the true form of $V_{TH}(H;T)$ levels off for $H \gg T$.
Thus, the potential using the quadratic approximation tends to overestimate
the height and width of the potential barrier of the vacuum transition for
small $|\lambda|$. From Figure \ref{fig:thrmpot} a--c, the quadratic
approximation of the thermal potential is good for $|\lambda| \sim 1 \gg
g_{\rm eff}^4 \sim 10^{-2}$, but clearly not for $|\lambda| \sim 10^{-2}$.

\begin{figure}
\begin{center}
\begin{tabular}{cc}
\includegraphics[width=.45\linewidth]{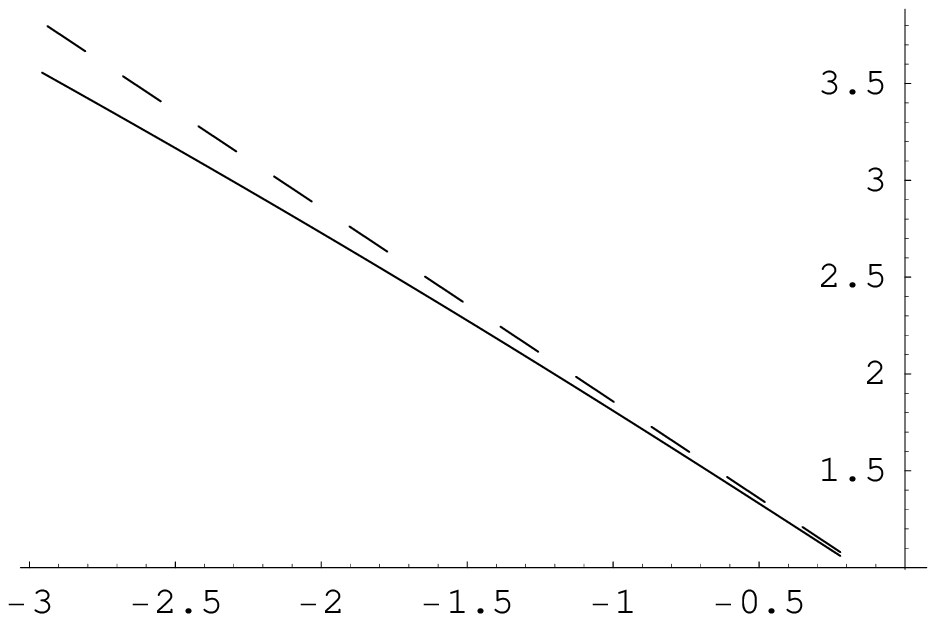} &
\includegraphics[width=.45\linewidth]{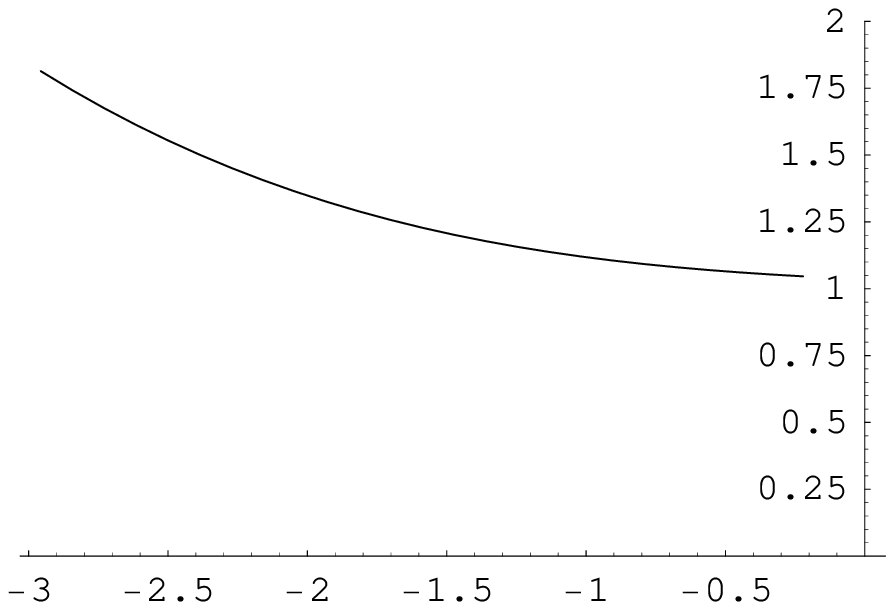}
\end{tabular}
\begin{picture}(0,0)(500,50)
%
%
\Text(450,-5)[lt]{$\log_{10}  |\lambda|$}
\Text(225,-5)[lt]{$\log_{10}  |\lambda|$}
\Text(470,125)[b]{\rm ratio}
\Text(250,120)[br]{$\log_{10} \left(\frac{S_3}{T}\right)$}
\end{picture}
\caption{\label{fig:highTexp} The bounce action $S_3/T$ with (dashed) and
without (solid) quadratic approximation of the thermal potential.}
\end{center}
\end{figure}

Since the quadratic approximation of the thermal potential $V_{TH}(H;T)$
overestimates the height and width of the potential barrier for small
$|\lambda|$, the bounce action $S_3 / T \simeq 6.015 \, \pi g_{\rm eff} /
|\lambda|$ based on the approximation is also an overestimation.
Figure \ref{fig:highTexp}a shows the bounce action calculated numerically
with the full form of the potential $V_{TH}(H;T)$ in (\ref{eq:WZT}) and 
with $V_{TH}^{(2)}(H; T)$ in (\ref{eq:thmass}).
Figure \ref{fig:highTexp}b shows the ratio between them.
As expected from the difference in the shape of the potential in
Figure \ref{fig:thrmpot}, the bounce action is overestimated in
(\ref{eq:SoverT})---by 20--30\% from Figure \ref{fig:highTexp}b
for $|\lambda| \sim 0.02\mbox{--}0.04$, the range of practical interest.
Thus, if the landscape prediction of the Higgs boson mass were calculated
upon the quadratic approximation, $\lambda_c(T)$ would have been 20--30\%
smaller, corresponding to $\lambda$ at the weak scale being 10\% smaller, 
and the prediction for $\overline{m}_H$ being 5\% lower.
It is not appropriate to use the quadratic approximation to calculate
$\lambda_c(T)$ and to obtain a precise landscape prediction.
All the calculations in section 4 use the potential (\ref{eq:WZT}), 
and the bounce solution and action were obtained and
calculated numerically. We used $m_H = 120$ GeV as the boundary 
condition of the 2-loop RG equations in order to calculate 
$g_Y$, $g_L$ and $h$ at high energy scale used in (\ref{eq:WZT}), 
but there is no impact on the result shown in
Figure~\ref{fig:mhprediction}, even if $m_H = 130$ GeV is used instead.

On the other hand, Figure \ref{fig:highTexp}a also tells us that
(\ref{eq:SoverT}) captures the qualitative aspects very well;
the numerically calculated bounce action (solid line) behaves almost
the same way as (\ref{eq:SoverT}). Thus, the expression (\ref{eq:SoverT})
as an approximation of the bounce action is still quite useful
when thinking of qualitative issues.

The biggest problem with the treatment of the thermal potential so far is
that Higgs loops have not been included in (\ref{eq:WZT}).\footnote{The contribution
from bottom quark loops is irrelevant to the calculation of the bounce
action so long as $h_b^4 \ll |\lambda|$.}  The contribution to $g_{\rm eff}^2$
has also been omitted; it would have been
\begin{equation}
g_{\rm eff.}^2 = \frac{1}{12} \left(\frac{3}{4}g_Y^2 + \frac{9}{4}g_L^2
			       + 3h^2 +	6 \lambda \right).
\end{equation}
It is technically quite involved to incorporate
the Higgs loop contribution to the thermal potential, when the high-temperature
expansion is no longer valid. This is beyond the scope of this paper.
Instead, let us try to get a feeling for the size of the possible effects of
the Higgs loop contribution,  using expressions in the quadratic
approximation, which seems to be qualitatively valid.
The $6 \lambda$ contribution in
$g_{\rm eff.}^2$ leads to a 5\% change in $g_{\rm eff.}$ for
$|\lambda|=0.03$.
A shift in $\lambda_c(T)$ of roughly  5\%  maintains the value of the
bounce action (\ref{eq:SoverT}), corresponding  to a 2\% change 
in the Higgs quartic coupling at the electroweak scale, and 1\% change 
in the Higgs mass prediction.
This estimate, however, heavily relies on the high-temperature expansion,
and thus may not be particularly accurate.


\begin{thebibliography}{99}
%
\bibitem{vacstab}
%
 N.~Cabibbo, L.~Maiani, G.~Parisi and R.~Petronzio,
  Nucl.\ Phys.\ B {\bf 158}, 295 (1979);
%
P.~Q.~Hung,
  Phys.\ Rev.\ Lett.\  {\bf 42}, 873 (1979);
%
 M.~Lindner,
  Z.\ Phys.\ C {\bf 31}, 295 (1986);
%
M.~Lindner, M.~Sher and H.~W.~Zaglauer,
  Phys.\ Lett.\ B {\bf 228}, 139 (1989).
%
\bibitem{vacstab2}
%
 M.~Sher,
  Phys.\ Rept.\  {\bf 179}, 273 (1989);
%
 B.~Schrempp and M.~Wimmer,
  Prog.\ Part.\ Nucl.\ Phys.\  {\bf 37}, 1 (1996)
  [arXiv:hep-ph/9606386].
%
\bibitem{vacstab3}
%
 M.~Sher,
  Phys.\ Lett.\ B {\bf 317}, 159 (1993)
  [Addendum-ibid.\ B {\bf 331}, 448 (1994)]
  [arXiv:hep-ph/9307342];
%
 G.~Altarelli and G.~Isidori,
  Phys.\ Lett.\ B {\bf 337}, 141 (1994);
%
 J.~A.~Casas, J.~R.~Espinosa and M.~Quiros,
  Phys.\ Lett.\ B {\bf 342}, 171 (1995)
  [arXiv:hep-ph/9409458];
%
 T.~Hambye and K.~Riesselmann,
  Phys.\ Rev.\ D {\bf 55}, 7255 (1997)
  [arXiv:hep-ph/9610272].
%
\bibitem{Strumia}
%
 G.~Isidori, G.~Ridolfi and A.~Strumia,
  Nucl.\ Phys.\ B {\bf 609}, 387 (2001)
  [arXiv:hep-ph/0104016].
%
\bibitem{mHdata}
%
The LEP Collaborations, the LEP Electroweak Working Group,
and the SLD Electroweak and Heavy Flavour Groups,
arXiv:hep-ex/0509008,
as updated on http://www.cern.ch/LEPEWWG
%
\bibitem{LHP}
%
  R.~Barbieri and A.~Strumia,
  arXiv:hep-ph/0007265.
%
\bibitem{Hoyle}
%
F. Hoyle, ``{\it Galaxies, nuclei and quasars},'' Heinemann, London, 1965, 
p 146 and p 159.
%
\bibitem{Agrawal:1997gf}
%
  V.~Agrawal, S.~M.~Barr, J.~F.~Donoghue and D.~Seckel,
  Phys.\ Rev.\ D {\bf 57}, 5480 (1998)
  [arXiv:hep-ph/9707380].
%
\bibitem{weinberg}
%
 S.~Weinberg,
  Phys.\ Rev.\ Lett.\  {\bf 59}, 2607 (1987); 
%
 H.~Martel, P.~R.~Shapiro and S.~Weinberg,
  Astrophys.\ J.\  {\bf 492}, 29 (1998)
  [arXiv:astro-ph/9701099].
%
\bibitem{stringlandscape}
%
 R.~Bousso and J.~Polchinski,
  JHEP {\bf 0006}, 006 (2000)
  [arXiv:hep-th/0004134]; 
%
 M.~R.~Douglas,
  JHEP {\bf 0305}, 046 (2003)
  [arXiv:hep-th/0303194]; 
%
  S.~Ashok and M.~R.~Douglas,
  JHEP {\bf 0401}, 060 (2004)
  [arXiv:hep-th/0307049]; 
%
 T.~Banks, M.~Dine and E.~Gorbatov,
  JHEP {\bf 0408}, 058 (2004)
  [arXiv:hep-th/0309170]; 
%
 F.~Denef and M.~R.~Douglas,
  JHEP {\bf 0405}, 072 (2004)
  [arXiv:hep-th/0404116]; 
%
 A.~Giryavets, S.~Kachru and P.~K.~Tripathy,
  JHEP {\bf 0408}, 002 (2004)
  [arXiv:hep-th/0404243]; 
%
  O.~DeWolfe, A.~Giryavets, S.~Kachru and W.~Taylor,
  JHEP {\bf 0502}, 037 (2005)
  [arXiv:hep-th/0411061].
%
\bibitem{FN}
%
C.~D.~Froggatt and H.~B.~Nielsen,
  Phys.\ Lett.\ B {\bf 368}, 96 (1996)
  [arXiv:hep-ph/9511371].
%
\bibitem{FNT}
%
C.~D.~Froggatt, H.~B.~Nielsen and Y.~Takanishi,
  Phys.\ Rev.\ D {\bf 64}, 113014 (2001)
  [arXiv:hep-ph/0104161]; 
%
\bibitem{LW}
%
K.~M.~Lee and E.~J.~Weinberg,
  Nucl.\ Phys.\ B {\bf 267}, 181 (1986).
%
\bibitem{GW}
%
A.~H.~Guth and E.~J.~Weinberg,
  Phys.\ Rev.\ D {\bf 23}, 876 (1981);
%
 A.~D.~Linde,
  Nucl.\ Phys.\ B {\bf 216}, 421 (1983)
  [Erratum-ibid.\ B {\bf 223}, 544 (1983)].
%
\bibitem{PDG06}
%
W.-M. Yao et al., J. Phys. G 33, 1 (2006) [Particle Data Group]. 
%
\bibitem{RGE}
%
C.~Ford, D.~R.~T.~Jones, P.~W.~Stephenson and M.~B.~Einhorn,
  Nucl.\ Phys.\ B {\bf 395}, 17 (1993)
  [arXiv:hep-lat/9210033].
%
\bibitem{threshold1}
%
A.~Sirlin and R.~Zucchini,
  Nucl.\ Phys.\ B {\bf 266}, 389 (1986).
%
\bibitem{Sher}
%
M.~Sher in \cite{vacstab3}.
%
\bibitem{threshold3}
%
R.~Hempfling and B.~A.~Kniehl,
  Phys.\ Rev.\ D {\bf 51}, 1386 (1995)
  [arXiv:hep-ph/9408313].
%
\bibitem{Hambaye}
%
 T.~Hambye and K.~Riesselmann in \cite{vacstab3}.
%
\bibitem{ADK}
%
  N.~Arkani-Hamed, S.~Dimopoulos and S.~Kachru,
  arXiv:hep-th/0501082.
%
\bibitem{Anderson}
%
G.~W.~Anderson,
  Phys.\ Lett.\ B {\bf 243}, 265 (1990).
%
\bibitem{AV}
%
 P.~Arnold and S.~Vokos,
  Phys.\ Rev.\ D {\bf 44}, 3620 (1991).
%
\bibitem{EQ}
%
J.~R.~Espinosa and M.~Quiros,
  Phys.\ Lett.\ B {\bf 353}, 257 (1995)
  [arXiv:hep-ph/9504241].
%
\bibitem{KT}
%
E.~Kolb, M.~Turner, {\it ``The Early Universe},'' 
Addison--Wesley, 1990.
%
\bibitem{Harnik:2006vj}
  R.~Harnik, G.~D.~Kribs and G.~Perez,
  arXiv:hep-ph/0604027.
%
\bibitem{Donoghue:2005cf}
  J.~F.~Donoghue, K.~Dutta and A.~Ross,
  Phys.\ Rev.\ D {\bf 73}, 113002 (2006)
  [arXiv:hep-ph/0511219].
%
\bibitem{Hall:1999sn}
  L.~J.~Hall, H.~Murayama and N.~Weiner,
  Phys.\ Rev.\ Lett.\  {\bf 84}, 2572 (2000)
  [arXiv:hep-ph/9911341].
%
\bibitem{Buch}
%
  W.~Buchmuller, P.~Di Bari and M.~Plumacher,
  Annals Phys.\  {\bf 315}, 305 (2005)
  [arXiv:hep-ph/0401240];
%
W.~Buchmuller, R.~D.~Peccei and T.~Yanagida,
  Ann.\ Rev.\ Nucl.\ Part.\ Sci.\  {\bf 55}, 311 (2005)
  [arXiv:hep-ph/0502169].
%
\bibitem{TR}
%
 M.~Tegmark and M.~J.~Rees,
  Astrophys.\ J.\  {\bf 499}, 526 (1998)
  [arXiv:astro-ph/9709058]; 
%
 M.~Tegmark, A.~Aguirre, M.~Rees and F.~Wilczek,
  Phys.\ Rev.\ D {\bf 73}, 023505 (2006)
  [arXiv:astro-ph/0511774].
%
\bibitem{barrowetal}
%
P.C.W Davies, {\it Journal of Physics} {\bf A5} 1296 (1972); 
L. Okamoto and C. Pask {\it Ann. Phys.} {\bf 68} 18 (1971);
J. Barrow and F. Tipler, {\it The Anthropic Cosmological Principle}, 
Clarendon Press, Oxford, 1986. 
%
\bibitem{Aguirre}
%
  A.~Aguirre,
   arXiv:astro-ph/0506519.
%
\bibitem{Wil}
A. Linde, Phys. Lett. B201, 437 (1988); F. Wilczek, hep-ph/0408167.
%
\end{thebibliography}
\end{document}